\documentclass[twocolumn]{aastex62}

\usepackage{placeins}

\newcommand*{\Cratio}{^{12}\mathrm{C}/^{13}\mathrm{C}}

\begin{document}
\title{Identifying Multiple Populations in M71 using CN}

\author{Jeffrey M. Gerber}
\affiliation{Department of Astronomy, Indiana University Bloomington, Swain West 319, 727 East 3rd Street, Bloomington, IN 47405-7105, USA}

\author{Eileen D. Friel}
\affiliation{Department of Astronomy, Indiana University Bloomington, Swain West 319, 727 East 3rd Street, Bloomington, IN 47405-7105, USA}
\author{Enrico Vesperini}
\affiliation{Department of Astronomy, Indiana University Bloomington, Swain West 319, 727 East 3rd Street, Bloomington, IN 47405-7105, USA}

\begin{abstract}
We have observed the CN features at $\sim$3800 ~\AA~ and 4120~ \AA~ as well as the CH band at $\sim$4300~\AA~ for 145 evolved stars in the Galactic globular cluster M71 using the multi-object spectrograph, Hydra, on the WIYN-3.5m telescope. We use these measurements to create two $\delta$CN indices finding that both distributions are best fit by two populations, a CN-enhanced and CN-normal. We find that 42 $\pm$ 4\% of the RGB stars in our sample are CN-enhanced. The percentage of CN-enhanced is 40 $\pm$ 13\% for the AGB and 33 $\pm$ 9\% for the HB stars, which suggests there are no missing second generation stars at these stages of stellar evolution. The two generations also separate in magnitude and color on the HB, which allows us to find the difference in He abundance between the two populations by fitting appropriate ZAHBs. The broad range of distances from the cluster's center covered by our sample allows us to study the dependence of the ratio of the number of first to second population stars on the distance from the cluster's center, and we find that this ratio does not vary radially and that the two populations are spatially mixed. Finally, we compare our identification of multiple populations with the classification based on the Na-O anti-correlation and the HST UV photometry, and we find good agreement with both methods.
\end{abstract}


\section{Introduction}
Once thought to be simple stellar populations, globular clusters (GCs) are now known to have much more complicated abundance patterns indicating the presence of multiple stellar populations. Studies of light elements have now revealed anti-correlations in C-N, Na-O, and Mg-Al indicative of multiple epochs of star formation taking place within GCs \citep[see e.g.][and references therein]{kraft94,gratton12}. Photometric surveys searching for multiple populations in Galactic GCs using HST UV photometry have also found that these multiple populations exist in every Galactic GC observed \citep[see e.g.][]{piotto15,milone17}. These studies show that GCs can no longer be considered as simple stellar populations. 

Multiple theories have been proposed to explain the formation of these multiple populations, and the progenitors that would have supplied the material to pollute the intracluster medium with the abundance patterns observed \citep[see e.g.][and references therein]{gratton12}. The possible sources of processed gas suggested in the literature include massive asymptotic giant branch (AGB) stars \citep[see e.g.][]{ventura01, dercole08, dercole12, dantona16}, fast-rotating massive stars \citep{prantzos06, decressin07b}, massive binary stars \citep{demink09}, massive stars \citep{elmegreen17}, and super-massive stars \citep{denissenkov14, gieles18}. However, no consensus has been reached on any model for the formation history of multiple populations and most questions concerning the origin of the observed abundance patterns are still unanswered.

The light element abundance patterns of the evolved, red giant branch (RGB) stars in GCs are complicated even further by a secondary mixing phenomenon that brings CN(O)-cycle material from the H-burning layer up to the surface of these stars. Secondary mixing begins in low mass stars (0.5 - 2.0 M$_\odot$) once they have passed the evolutionary stutter of the luminosity function bump (LFB). After passing this point on the RGB, the influx of CN(O)-cycle material being brought to the surface causes the surface abundance of carbon to decrease while nitrogen increases. Multiple theories to explain this non-canonical mixing have been presented and include thermohaline mixing, rotation, and magnetic fields, with thermohaline mixing often being favored \citep[see e.g.,][and references therein]{kippenhahn80,eggleton2006,charbonnel,denissenkov11,traxler11,wachlin11,brown13,henkel17}.

Due to these complications, large sample sizes ($\sim$100 or more) are needed to study the light element abundance patterns of evolved stars in GCs. Especially when studying carbon and nitrogen, large numbers are needed to separate the effects of multiple populations and secondary mixing in causing inhomogeneities in abundance patterns; both effects cause an anti-correlation in carbon and nitrogen since both involve some form of ``pollution" from CN(O)-cycled material.

In this paper, we continue our previous work on multiple populations in GCs \citep[][hereafter referred to as ``G18"]{gerber18} with a sample size of approximately 150 stars in M71; a relatively high metallicity Galactic GC with [Fe/H] = -0.78 dex (\citealt{harris}; 2010 edition). M71 is an interesting cluster to study for a number of reasons. It has a metallicity similar to the well-studied southern GC 47 Tuc, which allows for comparisons to be drawn between the two. M71 also has a handful of previous measurements of the CN and CH bands (although with much smaller sample sizes), which means it is the perfect candidate for a high metallicity comparison for our method of determining multiple populations outlined in G18. Na and O measurements and HST UV photometry also exist for this cluster, which allows us to make comparisons across methods of identifying multiple populations.

By studying the CN and CH bands in the near-UV, low-resolution spectra for 101 RGB, 15 AGB, and 27 HB stars in this cluster, we are able to classify stars into first and second generation across multiple evolutionary stages. We also focus on stars with Na-O measurements to compare classification methods, and see if stars are classified into different populations based on what method is used. Recent studies \citep[e.g.,][]{smith13,smith15a,smith15b,bobergm53} have revealed the presence of possible outliers classified differently depending on whether Na or CN is used. Our larger sample size allows us to see just how prevalent these outliers are, and determine if they are a statistically significant group of stars.

\section{Data}
\subsection{Observations and Target Selection} \label{sec::observations}
We obtained 171 combined spectra of stars within a 18' by 18' grid around the center of M71 between four observation runs in Aug. 2014, Jun. 2015, Jun. 2016, and Jun. 2017 using the Wisconsin-Indiana-Yale-NOAO (WIYN) \footnote{The WIYN Observatory is a joint facility of the University of Wisconsin-Madison, Indiana University, the National Optical Astronomy Observatory, the University of Missouri, and Purdue University} 3.5m Telescope and Hydra, a multi-object, fiber-fed bench spectrograph. The Bench Spectrograph was used with the ``600@10.1" grating, which resulted in spectra with a $\sim4.5$~\AA~pixel$^{-1}$ dispersion covering a range of $\sim2800$~\AA. The spectra taken during the 2014 run are centered on a wavelength of $4900$~\AA, while the spectra taken in subsequent runs are centered on a wavelength of $5100$~\AA. Ten different configurations of fibers were necessary to obtain our full sample size.

\begin{figure*}[htbp]
\centering
\includegraphics[trim = 0.4cm 0.4cm 0.4cm 0.4cm, scale=0.45, clip=True]{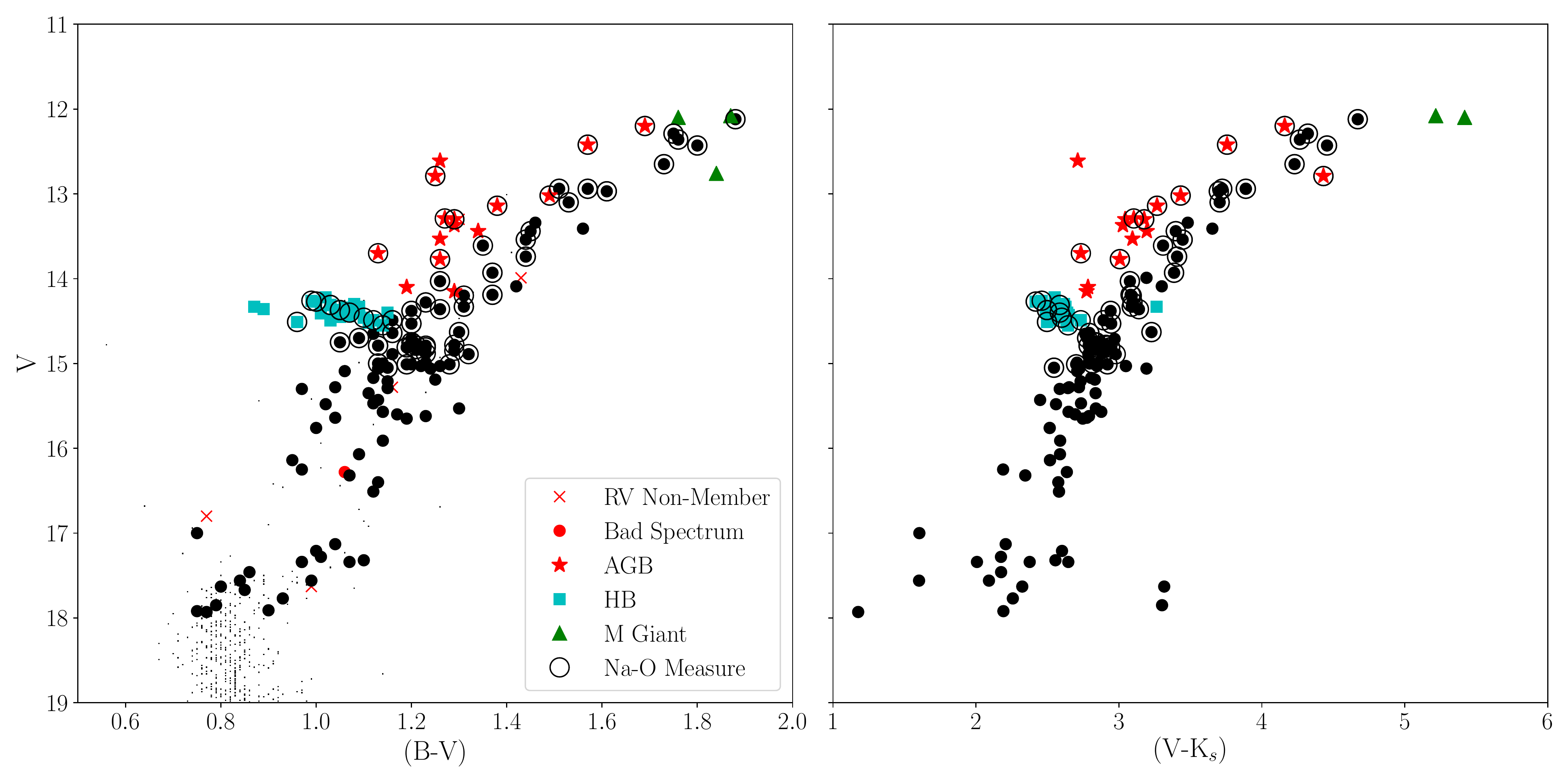}
\caption{\textbf{Left Panel:} CMD for M71 using B and V photometry from \citet{cudworth85}. Stars shown have an 85\% or higher probability of being cluster members based on their proper motions \citep{cudworth85}. Stars in our sample are indicated as large filled circles for RGB, red stars for AGB, and blue squares for HB. Three stars determined to be M giants are shown as green triangles. Stars with Na and/or O measurements are circled \citep{sneden94,ramirez02,carretta09a,carretta09b,cordero15}. \textbf{Right Panel:} CMD for M71 using V photometry from \citet{cudworth85} and K$_s$ photometry from 2MASS \citep{2mass}. Symbols are the same as the left panel.}
\label{fig::cmd}
\end{figure*}

We selected our target stars based on their location in the V vs. (B-V) CMD using photometry from \citet{cudworth85}. The final selection included stars from the tip of the RGB down to the MSTO, which occurs at a V $\sim$ 18, and is shown in Figure \ref{fig::cmd}. We only observed stars with an 85\% or greater likelihood of being cluster members based on proper motions determined by \citet{cudworth85}. We also chose to observe as many stars as possible with Na and/or O measurements by \citet{sneden94}, \citet{ramirez02}, \citet{carretta09a,carretta09b}, or \citet{cordero15}. Stars with measured Na and/or O abundances are indicated in Figure \ref{fig::cmd}. Our final selection covers $\sim$2.5 half-light radii from the cluster center, and allows us to study trends of the properties of multiple populations with both magnitude and radial distance from the cluster's center.

\subsection{Data Reduction and Radial Velocities}

We reduced the data with the IRAF\footnote{IRAF is distributed by the 
National Optical Astronomy Observatory, which is operated by the Association 
of Universities for Research in Astronomy (AURA) under cooperative agreement 
with the National Science Foundation.} package, \textit{dohydra} to perform bias subtraction, flat fielding, and dispersion correction with a CuAr comparison lamp spectrum. Final one-dimensional spectra were then produced using \textit{dohydra}, as well. We exposed configurations of brighter cluster stars (with $M_v \leq 0.5$) for a total integration time of $\sim$1 hr over 7-10 exposures and configurations of faint stars (with $M_V > 0.5$) for a total integration time of $\sim$2.5-3 hrs over 5-6 exposures to reduce the effects of cosmic rays and prevent saturation of the CCD. The processed spectra from each exposure were combined using the IRAF task \textit{scombine}. 

After reducing the spectra, radial velocities (RV) for each star were determined using the IRAF task \textit{fxcor} by cross-correlating with an RV standard star, HD 107328, observed during the 2016 and 2017 runs. For the 2014 and 2015 runs, RV measurements were made using the 2017 observations of the standard star. To check for consistency between standards used, we also made RV measurements using the 2016 observations of the same standard star and no systematic differences were found. All four runs showed further agreement by having similar medians, and stars with multiple observations across different runs were also found to have RV measurements that agreed within uncertainties and showed no systematic offsets between runs. The final RV measurements for each star are given in Table 1 along with positions and photometry from \citet{cudworth85}. Uncertainties on RV measurements made with this method at this resolution are around 15 km s$^{-1}$. We determined a median RV of -23 km s$^{-1}$ with a standard deviation of 6.5 km s$^{-1}$ for the cluster, which agrees well with what was found by high resolution studies of \citet{peterson86} (-22.1 $\pm$ 0.8 km s$^{-1}$), \citet{cohen01} (-21.7 $\pm$ 2.6 (s.t.d.) km s$^{-1}$), and \citet{kimmig15} (-23.1 $\pm$ 0.3 km s$^{-1}$). 

Final membership was determined using a similar method as in G18; stars with radial velocities falling outside of three standard deviations from the cluster median when calculated using the entire sample were deemed as non-members. Because our stars were selected based on a proper motion study, we do not have many non-members in our sample. Our sample includes 171 RV measurements with 5 stars being RV non-members shown in Figure \ref{fig::cmd} as red crosses. Of the remaining spectra, 20 were multiple measurements of stars across nights or observing runs to use to determine uncertainties and one was found to have a bad spectrum (too low S/N). This leaves the final number of likely member stars observed as 145, with 100 RGB, 15 AGB, 27 HB stars, and three stars determined to be M giants also shown in Figure \ref{fig::cmd}. 2MASS photometry was used to identify AGB and HB stars as the relatively high amount of extinction towards the direction of M71 made stars difficult to classify in the optical CMD; the RGB, AGB, and HB sequences were much more clearly defined in the IR photometry as is shown in the right panel of Figure \ref{fig::cmd}.

\section{Analysis - CN and CH Bands}
\subsection{Index Definitions}
Following our previous work (G18), we used the CN index at $\sim$3885 \AA, S(3839), and the commonly used CH index at $\sim$4300 \AA~as defined by \citet{bands}. We were also able to use the CN index at $\sim$4140 \AA, CN(4216), \citep{friel87} due to the relatively high metallicity of M71 stars. The higher metallicity allows the redder CN band, which is the weaker and broader of the two bands, to be strong enough to give significant measurements and show a difference between the two populations. We used the CN(4216) index as defined by \citet{friel87} rather than the commonly used S(4142) index from \citet{s4142} because CN(4216) is a double sided index and therefore, less likely to be affected by any problems with the continuum levels of the spectra. Our final index definitions are as follows:

\begin{equation}
S(3839) = - 2.5log\frac{F_{3861-3884}}{F_{3894-3910}}
\end{equation}
\begin{equation}
CN(4216) = - 2.5log\frac{F_{4144-4177.75}}{0.5F_{4082-4118.25}+0.5F_{4246-4284.75}}
\end{equation}
\begin{equation}
CH(4300) = -2.5log\frac{F_{4285-4315}/30}{0.5(F_{4240-4280}/40) + 0.5(F_{4390-4460}/70)}
\end{equation}
\\
\subsection{Flux Calibration}
We followed the method used in G18 to flux calibrate the spectra using model spectra generated by the Synthetic Spectrum Generator (SSG) \citep[][and references therein]{ssg} using MARCS model atmospheres \citep{gustafsson08}. We used a metallicity of [Fe/H] = -0.78 (\citealt{harris} (2010 edition)). As in G18, effective temperatures for the model spectra were calculated from the V-K colors of the stars based on the method by \citet{alonso99, alonso01} with V magnitudes from \citet{cudworth85} and K magnitudes from 2MASS \citep{2mass}. K magnitudes were corrected to the TCS system following the method by \citet{johnson05}. If a star did not have a K magnitude, the relation for effective temperature based on B-V color was used \citep{alonso99}.

M71 is located at a low Galactic latitude, and therefore, suffers from more extinction than most other GCs. Measurements for the reddening of the cluster range from E(B-V) = 0.21 to 0.32 \citep[][and references therein]{brasseur10}. We note that differences also seem to exist depending on what photometric filters were used to derive the E(B-V) for the cluster. For example, studies matching models to CMDs using filters in the optical region tend to find a higher E(B-V) of 0.28 \citep{grundahl2002,morrison16}, while studies matching CMDs in the infrared tend to find a lower E(B-V) of 0.22 \citep{brasseur10}. Because of these discrepancies and the fact that we had to use a mixture of infrared and optical colors to determine accurate temperatures, we selected an E(B-V) between these values that allowed for the highest consistency between effective temperatures calculated with B-V and V-K colors. This method ensured that all of our stars were on the same temperature scale whether they had accurate IR photometry or not. 
Recent photometric studies of M71 have also indicated that there is differential extinction across the cluster \citep{morrison16}, so we used the reddening map from \citet{morrison16} to correct for differential extinction, which can cause changes in the E(B-V) value up to 0.08.

Surface gravities were calculated using the bolometric corrections given by \citet{alonso99} and are listed with the effective temperature for each star in Table 1. Absolute magnitudes were calculated for each star using an apparent distance modulus of (m-M)$_V$ = 13.728 from \citet{morrison16} and an extinction value for each star calculated using the reddening from \citet{morrison16}. These are also given in Table 1.

\subsection{Band Measurements} \label{sec::bands}
\begin{figure}[htbp]
\centering
\includegraphics[trim = 0.4cm 0.4cm 0.4cm 0.4cm, scale=0.4, clip=True]{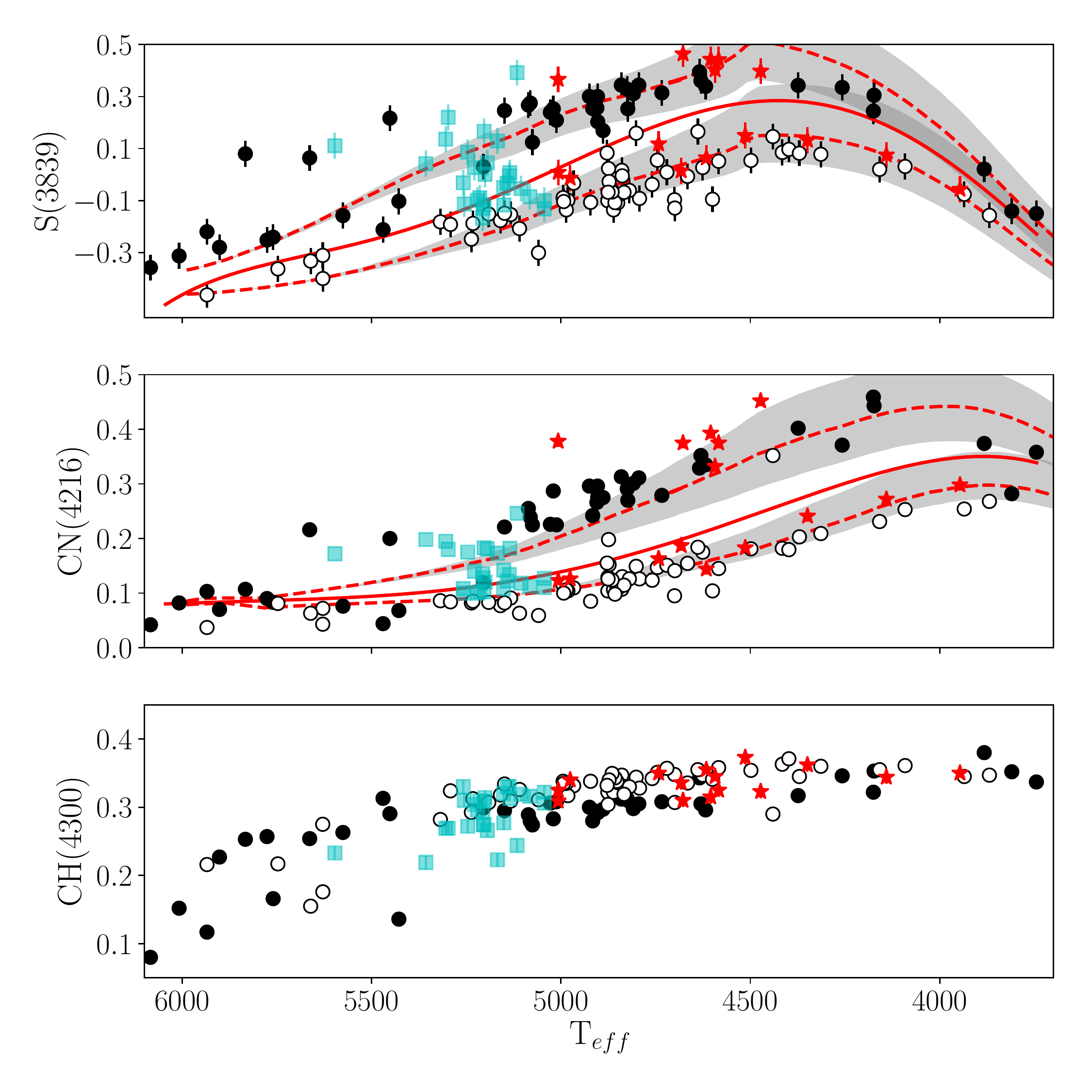}
\caption{S(3839), CN(4216), and CH indices plotted against effective temperature, top, middle, and bottom panel, respectively. Points are indicated as in Figure \ref{fig::cmd} with the exception that CN-enhanced RGB stars are indicated as filled circles and CN-normal RGB stars are indicated as open circles. Error bars showing the estimated uncertainty are shown in each panel. For CN(4216) and CH, the estimated uncertainty is smaller than the points. Solid red lines in the top two panels are the fiducial lines used to create a $\delta$S(3839) and a $\delta$CN(4216) index. Dashed red lines indicate fiducials with a given C and N abundance; [C/Fe] = -0.3 dex, [N/Fe] = 1.5 dex for the CN-enhanced and [C/Fe] = 0.0, [N/Fe] = 0.6 dex for the CN-normal. The shaded, grey areas around these fiducials represent the uncertainty from the range in [O/Fe] abundances for these stars as discussed in Section \ref{sec::bands}.}
\label{fig::bands}
\end{figure}

We measured CN and CH band strengths of the flux calibrated spectra and plotted the measurements vs. effective temperature as shown in Figure \ref{fig::bands}. Uncertainties on our band strength measurements were calculated as in G18. The standard deviation of band strengths for stars that were measured across multiple nights or runs was determined for faint (M$_\mathrm{V} \geq$ 1.0) and bright (M$_\mathrm{V} <$ 1.0) stars. The uncertainty was roughly the same for both magnitude groups, so a combined median standard deviation was taken for all stars as representative of the uncertainty in our measurements. This method gives uncertainties of 0.1 for the S(3939) index, 0.05 for the CN(4216) index, and 0.05 for the CH(4300) index.

In G18, we showed how the CN and CH band strengths are affected by effective temperature and surface gravity by plotting them vs. absolute magnitude (luminosity). For this paper, we have chosen to plot band strengths vs. effective temperature instead, so that stars in different evolutionary stages (RGB, AGB, and HB) all follow similar trends. By plotting vs. effective temperature, we help offset effects other than abundances that cause changes in band strengths between different evolutionary stages at a given magnitude such as AGB and HB stars being hotter than RGB stars.

Looking at the S(3839) band in the top panel of Figure \ref{fig::bands}, there is a clear spread in CN band strength that would indicate the presence of multiple populations, but populations do not separate themselves very clearly as in M10 (G18). However, the middle panel of Figure \ref{fig::bands} shows two populations clearly separated in CN(4216) strength similar to what was seen by \citet{briley01b} in DDO photometry of RGB stars in M71. While this separation seems certain enough to classify stars in M71 by eye alone, we still needed to create an objective classification method for two reasons. The first is that the CN(4216) band becomes too weak at the higher effective temperatures of the faintest stars to be sensitive to abundance differences, and therefore, cannot be used to classify the faintest stars in our sample (M$_\mathrm{V} \geq$ 3.0; T$_{eff} \geq$ 5300 K). Second, since the CN(4216) band cannot be used to classify the faintest stars we must rely on the S(3839) band, but because the populations do not separate as clearly in the S(3839) band, an objective classification method is required.

To classify populations objectively, we followed the same method as that in G18 by modeling the atmospheric effects on the band strengths and creating a $\delta$CN index. Atmospheric effects were modeled by generating synthetic spectra with the SSG. Effective temperatures and surface gravities were taken from a 12 Gyr PARSEC isochrone \citep{marigo17}. An age of 12 Gyr was chosen based on the age for the cluster found by \citet{dicecco15}. 

We based carbon and nitrogen abundances for these spectra on the studies of \citet{briley01a, briley01b, briley04b}. \citet{briley01b} found two populations of RGB stars in DDO photometry of the CN(4216) band and were able to match sets of carbon and nitrogen abundances to each population. The CN-normal population approximately matched a [C/Fe] of 0.0 dex and a [N/Fe] of 0.4 dex, while the CN-enhanced population approximately matched a [C/Fe] of -0.3 dex and a [N/Fe] of 1.4 dex. The same sets of abundances were then found to match S(3839) and CH band measurements of main sequence stars by \citet{briley01a}. These bands were then used to make abundance measurements for individual stars by \citet{briley04b}, which also agreed with those previously measured for each population. 

We used these carbon and nitrogen abundances as starting points to find a carbon and nitrogen abundance that would match each population in the S(3839) and CN(4216) band strengths. Since the work by \citet{briley01a,briley01b}, and \citet{briley04b}, oxygen abundances have been measured for large samples of stars in M71 \citep{carretta09a,carretta09b,cordero15}, so some adjustments had to be made to these abundances. The measured oxygen abundances were slightly higher than those assumed by \citet{briley01a, briley01b}, and \citet{briley04b}, which means that the nitrogen abundances had to be raised to match CN band strengths. We find that a fiducial created with [C/Fe] = -0.3 dex and [N/Fe] = 1.5 dex approximately matches the CN-enhanced population, and a fiducial with [C/Fe] = 0.0 dex and [N/Fe] = 0.6 dex matches the CN-normal population. We used an [O/Fe] = 0.3 dex and [O/Fe] = 0.41 dex for the CN-enhanced and CN-normal populations, respectively, based on the measurements from various sources \citep{sneden94,ramirez02,carretta09a,carretta09b,cordero15}. We also had to take into account that the surface $\Cratio$ changes in GC RGB stars as they climb the giant branch due to secondary mixing. We followed the same method as \citet{simpson17} and \citet{kirby15} who also studied RGB stars in GCs and adjusted the $\Cratio$ as a function of surface gravity based on measurements shown in Figure 4 in \citet{keller01}. 

When creating fiducial lines in the S(3839) vs. M$_\mathrm{V}$ for M10 in G18, we had to take the change in abundance as a function of magnitude into account. We determined this was not necessary for M71 as the efficiency of extra mixing will be lower for a higher metallicity cluster. \citet{smith07} showed that the variation of [C/Fe] as a function of magnitude for some of the brightest stars in M71 is less than 0.2 dex. The final fiducial lines for each population are plotted in Figure \ref{fig::bands} as red dashed lines with each line matching the corresponding population relatively well in both CN band strengths. These fiducial lines also have a certain amount of uncertainty caused by the spreads in oxygen found in each population \citep{sneden94,ramirez02,carretta09a,carretta09b,cordero15}. The [O/Fe] for the CN-normal population ranges from 0.1 to 0.6 dex, and 0.0 to 0.6 dex for the CN-enhanced population. The shaded areas around each population fiducial shown in Figure \ref{fig::bands} represent the uncertainties caused by spreads between these values.

For the fiducial used to separate the two populations, we took the average carbon, nitrogen, and oxygen abundances from these CN-enhanced and CN-normal fiducials (i.e., [C/Fe] = -0.15, [N/Fe] = 1.0, [O/Fe] = 0.355 dex). Synthetic spectra were then used to create a fiducial average band strength for the S(3839) band and the CN(4216) band, which are plotted in Figure \ref{fig::bands} as solid red lines. A $\delta$S(3839) index and a $\delta$CN(4216) index were generated by subtracting the S(3839) and CN(4216) band strengths from their given average fiducials (see G18 for more details).

RGB stars with a positive $\delta$S(3839) index are considered CN-enhanced stars and are represented in Figure \ref{fig::bands} as closed circles, while stars with a negative $\delta$S(3839) index are CN-normal and shown as open circles. We keep this designation for all three panels in Figure \ref{fig::bands}. In the middle panel of Figure \ref{fig::bands}, the two populations classified by the S(3839) index fall exactly where one would expect for the stars cool enough (T$_{eff} \leq$ 5300 K) to have CN(4216) band strengths sensitive to changes in abundance except for two stars. One of these stars is enhanced in $\delta$S(3839), but not in $\delta$CN(4216). This difference can be explained by the fact that it is one of the brightest stars in the sample, and the S(3839) band is not very sensitive to differences in abundances at that magnitude. The other star is one that is enhanced in CN(4216) and not in S(3839), which is due to the spectrum for this star having lower S/N than the others. Since our sensitivity decreases towards bluer wavelengths, the bluer S(3839) band was more heavily affected by low S/N and resulted in too low a measurement. While the CN(4216) index was not used directly for classification, it provides confidence in our classification method because both bands seem to give consistent assignment to populations of stars in the cluster.

\begin{deluxetable*}{ccccccccccccccccccccc}
\tabletypesize{\tiny}
\tablecolumns{21}
\tablewidth{0pt}
\tablecaption{Stars Measured in M71\label{tab::alldata}}
\label{tab::all}
\tablehead{\\
   \colhead{ID$^{1}$} &          \colhead{RA$^{1}$} &        \colhead{Dec$^{1}$} &     \colhead{RV$_{hel}$} &      \colhead{V$^{1}$} &    \colhead{(B-V)$^{1}$} &    \colhead{M$_{\mathrm{V}}^{2}$} &          \colhead{ID$_{2MASS}$} &    \colhead{T$_{eff}$} &      \colhead{logg} &  \colhead{S(3839)} &  \colhead{CN(4216)} &  \colhead{CH(4300)} &  \colhead{$\delta$S(3839)} &  \colhead{$\delta$CN(4216)} &  \colhead{[Na/Fe]} &  \colhead{[O/Fe]} & \colhead{Na Source$^{3}$} & \colhead{O Source$^{3}$} & \colhead{Branch$^{4}$} & \colhead{Memb}\\ & \colhead{(J2000)} & \colhead{(J2000)} & \colhead{(km s$^{-1}$)} & \colhead{mag} & \colhead{mag} & \colhead{mag} & (K) & & & & & & & & &}
\rotate
\startdata
    2 &  298.406125 &  18.749889 &  -26 &  12.94 &  1.57 & -0.85 &  19533747+1844596 &  4091 &  1.38 &    0.031 &     0.212 &     0.361 &           -0.123 &            -0.084 &    0.130 &   0.360 &        Cord &       Cord &    RGB &    y \\
    3 &  298.455875 &  18.740778 &  -23 &  12.08 &  1.87 & -1.62 &  19534941+1844269 &  3601 &  0.41 &   \nodata &     \nodata &     \nodata &            \nodata &             \nodata &   \nodata &  \nodata &     \nodata &    \nodata &      M &    y \\
    5 &  298.464542 &  18.801667 &  -22 &  12.29 &  1.75 & -1.44 &  19535150+1848059 &  3869 &  0.93 &   -0.157 &     0.202 &     0.347 &           -0.081 &            -0.081 &    0.320 &   0.520 &        Cord &       Cord &    RGB &    y \\
    6 &  298.444167 &  18.778083 &  -22 &  13.34 &  1.46 & -0.39 &  19534660+1846411 &  4314 &  1.74 &    0.078 &     0.180 &     0.360 &           -0.201 &            -0.076 &   \nodata &  \nodata &     \nodata &    \nodata &    RGB &    y \\
    9 &  298.401125 &  18.824889 &  -25 &  13.41 &  1.56 & -0.44 &  19533626+1849297 &  4257 &  1.67 &    0.335 &     0.327 &     0.346 &            0.076 &             0.071 &   \nodata &  \nodata &     \nodata &    \nodata &    RGB &    y \\
   14 &  298.362875 &  18.766361 &   30 &  13.30 &  1.30 & -0.52 &  19532709+1845588 &  4727 &  1.94 &    0.380 &     0.265 &     0.367 &            0.162 &             0.130 &   \nodata &  \nodata &     \nodata &    \nodata &    AGB &    n \\
   15 &  298.371875 &  18.766222 &  -23 &  13.74 &  1.44 & -0.08 &  19532925+1845584 &  4416 &  1.93 &    0.085 &     0.155 &     0.363 &           -0.212 &            -0.075 &    0.228 &   0.281 &      Car09a &     Car09a &    RGB &    y \\
   22 &  298.413250 &  18.734167 &  -24 &  13.93 &  1.37 &  0.17 &  19533918+1844030 &  4398 &  2.02 &    0.097 &     0.154 &     0.371 &           -0.198 &            -0.083 &    0.150 &   0.560 &        Cord &       Cord &    RGB &    y \\
   23 &  298.416125 &  18.731389 &  -22 &  12.94 &  1.51 & -0.82 &  19533986+1843530 &  4173 &  1.46 &    0.305 &     0.372 &     0.353 &            0.091 &             0.122 &    0.340 &   0.050 &        Cord &       Cord &    RGB &    y \\
   27 &  298.510625 &  18.762056 &  -23 &  14.10 &  1.19 &  0.31 &  19540255+1845434 &  4976 &  2.40 &   -0.014 &     0.106 &     0.340 &           -0.077 &            -0.006 &   \nodata &  \nodata &     \nodata &    \nodata &    AGB &    y \\
   32 &  298.503083 &  18.818333 &  -19 &  12.61 &  1.26 & -1.12 &  19540075+1849061 &  5007 &  1.84 &    0.366 &     0.332 &     0.309 &            0.325 &             0.250 &   \nodata &  \nodata &     \nodata &    \nodata &    AGB &    y \\
   37 &  298.406583 &  18.791250 &  -31 &  12.65 &  1.73 & -1.17 &  19533757+1847286 &  3935 &  1.11 &   -0.076 &     0.214 &     0.345 &           -0.077 &            -0.098 &    0.509 &   0.396 &      Car09b &     Car09b &    RGB &    y \\
   38 &  298.409042 &  18.780472 &   26 &  13.99 &  1.43 &  0.23 &  19533817+1846497 &  4551 &  2.14 &    0.393 &     0.443 &     0.360 &            0.109 &             0.278 &   \nodata &  \nodata &     \nodata &    \nodata &    RGB &    n \\
   39 &  298.394625 &  18.772611 &  -21 &  14.09 &  1.42 &  0.27 &  19533470+1846213 &  4498 &  2.12 &    0.054 &     0.154 &     0.354 &           -0.240 &            -0.053 &   \nodata &  \nodata &     \nodata &    \nodata &    RGB &    y \\
   44 &  298.468125 &  18.748389 &  -29 &  14.03 &  1.26 &  0.30 &  19535236+1844542 &  4630 &  2.21 &    0.361 &     0.326 &     0.305 &            0.101 &             0.152 &    0.450 &   0.160 &        Cord &       Cord &    RGB &    y \\
   45 &  298.479417 &  18.759583 &  -22 &  13.77 &  1.26 & -0.02 &  19535507+1845344 &  4742 &  2.15 &    0.118 &     0.131 &     0.350 &           -0.092 &            -0.011 &    0.390 &   0.530 &        Cord &       Cord &    AGB &    y \\
   46 &  298.471833 &  18.779750 &  -19 &  12.20 &  1.69 & -1.56 &  19535325+1846471 &  3946 &  0.96 &   -0.057 &     0.250 &     0.350 &           -0.071 &            -0.054 &    0.180 &   0.330 &        Cord &       Cord &    AGB &    y \\
   47 &  298.474958 &  18.784250 &  -15 &  12.76 &  1.84 & -1.00 &  19535399+1847034 &  3485 &  0.19 &   \nodata &     \nodata &     \nodata &            \nodata &             \nodata &   \nodata &  \nodata &     \nodata &    \nodata &      M &    y \\
   48 &  298.483708 &  18.787972 &   16 &  12.10 &  1.76 & -1.63 &  19535610+1847167 &  3568 &  0.30 &   \nodata &     \nodata &     \nodata &            \nodata &             \nodata &   \nodata &  \nodata &     \nodata &    \nodata &      M &    y \\
   49 &  298.490167 &  18.799167 &  -26 &  13.44 &  1.45 & -0.29 &  19535764+1847570 &  4374 &  1.82 &    0.343 &     0.366 &     0.317 &            0.051 &             0.133 &    0.660 &   0.230 &        Cord &       Cord &    RGB &    y \\
   52 &  298.461000 &  18.818722 &  -25 &  12.97 &  1.61 & -0.76 &  19535064+1849075 &  4175 &  1.49 &    0.244 &     0.384 &     0.322 &            0.029 &             0.139 &    0.820 &   0.220 &        Cord &       Cord &    RGB &    y \\
   53 &  298.463333 &  18.808583 &  -23 &  14.36 &  1.26 &  0.63 &  19535120+1848308 &  4666 &  2.37 &   -0.007 &     0.121 &     0.336 &           -0.253 &            -0.037 &    0.200 &  \nodata &        Cord &       \nodata &    RGB &    y \\
   55 &  298.451167 &  18.807028 &  -18 &  13.14 &  1.38 & -0.59 &  19534828+1848253 &  4473 &  1.76 &    0.398 &     0.370 &     0.323 &            0.102 &             0.211 &    0.420 &   0.350 &        Cord &       Cord &    AGB &    y \\
   61 &  298.451125 &  18.800611 &  -21 &  12.36 &  1.76 & -1.34 &  19534827+1848021 &  3882 &  0.98 &    0.021 &     0.309 &     0.380 &            0.081 &             0.023 &    0.680 &   0.110 &        Cord &       Cord &    RGB &    y \\
   62 &  298.446458 &  18.800333 &  -20 &  13.44 &  1.34 & -0.26 &  19534715+1848012 &  4513 &  1.92 &    0.150 &     0.159 &     0.373 &           -0.142 &            -0.047 &   \nodata &  \nodata &     \nodata &    \nodata &    AGB &    y \\
   63 &  298.440625 &  18.798583 &  -14 &  12.43 &  1.80 & -1.27 &  19534575+1847547 &  3810 &  0.92 &   -0.141 &     0.217 &     0.352 &            0.008 &            -0.059 &    0.230 &   0.310 &        Cord &       Cord &    RGB &    y \\
   64 &  298.444000 &  18.795583 &  -26 &  13.53 &  1.26 & -0.17 &  19534656+1847441 &  4593 &  2.00 &    0.403 &     0.297 &     0.345 &            0.130 &             0.123 &   \nodata &  \nodata &     \nodata &    \nodata &    AGB &    y \\
   65 &  298.442292 &  18.790556 &  -19 &  13.10 &  1.53 & -0.60 &  19534615+1847261 &  4159 &  1.54 &    0.020 &     0.202 &     0.355 &           -0.185 &            -0.093 &    0.360 &   0.440 &        Cord &       Cord &    RGB &    y \\
   66 &  298.432208 &  18.792667 &  -28 &  13.54 &  1.44 & -0.25 &  19534373+1847336 &  4370 &  1.83 &    0.082 &     0.164 &     0.345 &           -0.210 &            -0.067 &    0.436 &   0.384 &      Car09a &     Car09a &    RGB &    y \\
   68 &  298.428667 &  18.777056 &  -26 &  13.70 &  1.13 & -0.06 &  19534288+1846374 &  5007 &  2.26 &    0.008 &     0.108 &     0.325 &           -0.033 &            -0.005 &    0.330 &  \nodata &        Cord &       \nodata &    AGB &    y \\
   71 &  298.420958 &  18.768222 &  -25 &  13.30 &  1.29 & -0.37 &  19534103+1846056 &  4584 &  1.92 &    0.442 &     0.342 &     0.325 &            0.166 &             0.163 &    0.400 &   0.310 &        Cord &       Cord &    AGB &    y \\
   72 &  298.428375 &  18.770278 &  -25 &  14.15 &  1.29 &  0.42 &  19534282+1846129 &  4616 &  2.25 &    0.064 &     0.126 &     0.355 &           -0.201 &            -0.059 &   \nodata &  \nodata &     \nodata &    \nodata &    AGB &    y \\
   75 &  298.432125 &  18.770944 &  -28 &  14.33 &  0.87 &  0.60 &  19534371+1846154 &  5597 &  2.77 &    0.111 &     0.140 &     0.233 &            0.403 &             0.081 &   \nodata &  \nodata &     \nodata &    \nodata &     HB &    y \\
   77 &  298.436708 &  18.763500 &  -22 &  14.26 &  0.99 &  0.53 &  19534480+1845486 &  5304 &  2.63 &    0.136 &     0.155 &     0.269 &            0.291 &             0.094 &    0.520 &  \nodata &        Cord &       \nodata &     HB &    y \\
   78 &  298.440083 &  18.763944 &  -21 &  14.37 &  1.05 &  0.64 &  19534561+1845502 &  5258 &  2.66 &   -0.032 &     0.076 &     0.330 &            0.096 &             0.004 &    0.230 &   0.560 &         Ram &        Ram &     HB &    y \\
\enddata
 \tablecomments{1. \citet{cudworth85}, 2. Assuming (m -- M)$_{v}$ = 13.728, 3. \citet{sneden94}, \citet{ramirez02}, \citet{carretta09a,carretta09b}, or \citet{cordero15}, 4. M indicates a star with spectral type M \\ The complete table is available online.}
\end{deluxetable*}

\section{Results and Discussion}
\subsection{Multiple CN Populations in M71} \label{sec::rgbpops}
\subsubsection{RGB stars}
\begin{figure}[htbp]
\centering
\includegraphics[trim = 0.2cm 0.4cm 0.4cm 0.4cm, scale=0.4, clip=True]{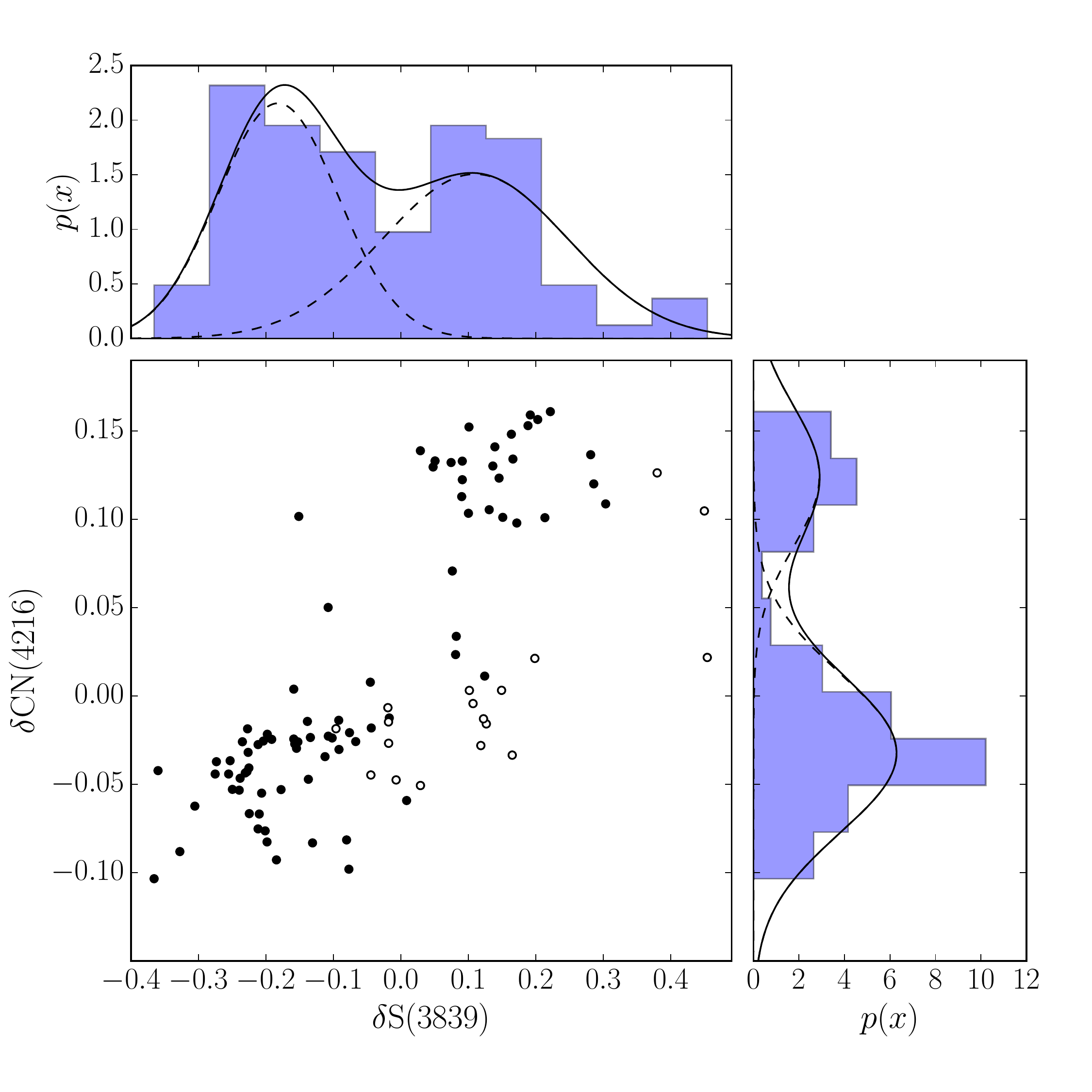}
\caption{$\delta$CN(4216) vs. $\delta$S(3839) for the RGB stars in our sample. RGB stars with T$_{eff} >$ 5300 K are indicated as open circles. Histograms for each axis are shown with fits based on Gaussian Mixture Models overplotted.}
\label{fig::deltacnrgb}
\end{figure}

Following G18 and as discussed in Section \ref{sec::bands}, we used the $\delta$S(3839) index to sort the RGB stars in M71 into CN-enhanced and CN-normal populations. The measured $\delta$CN(4216) vs. $\delta$S(3839) index for the RGB stars in our sample is shown in Figure \ref{fig::deltacnrgb}. In general, the two $\delta$CN bands correlate well with one another as expected from Figure \ref{fig::bands}. There is a small group of stars with high $\delta$S(3839) and low $\delta$CN(4216), which are the fainter stars in the sample. For these stars, the surface temperature becomes hot enough that the CN(4216) band is no longer sensitive to changes in abundance (T$_{eff} \geq$ 5300 K). 

The presence of at least two populations is evident from the histograms of $\delta$S(3839) and $\delta$CN(4216) along the x and y axes of Figure \ref{fig::deltacnrgb}. However, in order to test whether M71 hosts additional populations, as detected in a few Galactic clusters \citep[see e.g.,][]{carretta09a,carretta09b,carretta15,milone15,milone15b}, we have calculated the number of populations by applying a Gaussian mixture model (GMM) to the $\delta$S(3839) and $\delta$CN(4216) distributions. 
The Bayesian and Akaike information criteria (BIC and AIC, respectively) are used to quantify the success of fitting different models with different numbers of populations. The model with the number of populations that minimizes both criteria is accepted as the best fit. For both the $\delta$S(3839) and the $\delta$CN(4216) distributions, the BIC and AIC were minimized with two Gaussians (one for each population). The CN-enhanced population had an average $\delta$S(3839) of 0.18 and an average $\delta$CN(4216) of 0.125, while the CN-normal population had an average $\delta$S(3839) of -0.17 and an average $\delta$CN(4216) of -0.025. The final GMM fits to each distribution are plotted in Figure \ref{fig::deltacnrgb}.

Based on this information and our $\delta$S(3839) index, we find that 42 out of 100 or 42 $\pm$ 4\% of RGB stars are CN-enhanced and belong to the second generation, making M71 a first generation dominated cluster. 
Our uncertainty comes from propagating uncertainties based on counting statistics through the calculation of the ratio of second generation stars to the total stars in our sample. 
This ratio agrees with what is found using photometry centered on the S(3839) band by \citet{bowman17} who found roughly 45\% of stars belonging to the second generation. However, the result disagrees with that found in the Na-O distribution by \citet{carretta09a} and \citet{cordero15}, with the latter finding M71 to be dominated by Na-enhanced stars with 71\% belonging to the second generation (using [Na/Fe] = 0.2 dex as a separating value). We leave our discussion on the possible reason for this disagreement to a later section (Section \ref{sec::na-o}) where we compare to other methods of identifying multiple populations. 

\subsubsection{AGB Stars}
Recently, there has been some debate as to whether or not all populations found in GCs exist at all stages of stellar evolution after the RGB; specifically, an absence of the second generation on the AGB has been reported in clusters such as NGC 6257 \citep{campbell12,campbell13} and M4 \citep{maclean16}, while second generation AGB stars have been observed in other clusters, namely, 47 Tuc, M13, M5, M3, M2, and NGC 6397 \citep[][see also Section 4.1.1 of G18 for further discussion and details on multiple populations in AGB stars]{johnson15,garcia15,maclean18}.

\begin{figure}[htbp]
\centering
\includegraphics[trim = 0.2cm 0.4cm 0.4cm 0.4cm, scale=0.4, clip=True]{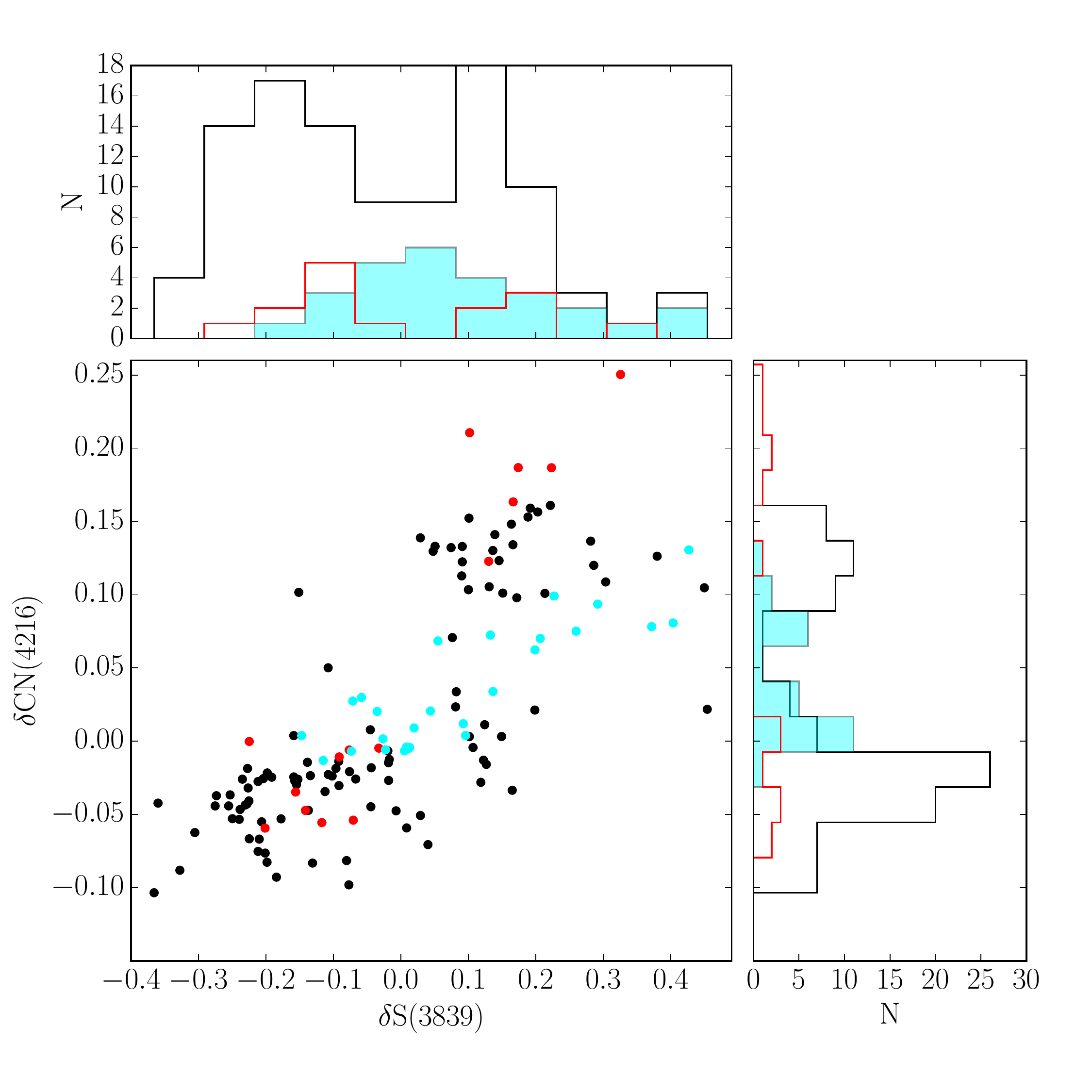}
\caption{$\delta$CN(4216) vs. $\delta$S(3839) for all stars in our sample. RGB stars are shown as black, AGB as red, and HB as blue points. Histograms for each axis are shown with color coding that matches the color coding for the points.}
\label{fig::deltacnall}
\end{figure}

In Figure \ref{fig::deltacnall}, we add the distributions of $\delta$S(3839) and $\delta$CN(4216) for the AGB and HB stars in M71 to Figure \ref{fig::deltacnrgb} as red and blue points, respectively. From Figure \ref{fig::deltacnall} it seems that the AGB stars roughly follow the distribution of the RGB stars with the only difference being a couple of stars with slightly higher $\delta$CN(4216) values. The AGB stars also separate very clearly into two populations in both the $\delta$S(3839) and $\delta$CN(4216) bands. However, as stated in G18, classifying AGB stars by comparing a $\delta$CN measurement to RGB stars without other considerations can lead to misclassification. This misclassification stems from the fact that while $\delta$CN can be used as a proxy for the underlying N abundance, the band is also dependent on other abundances that should be considered, particularly the C abundance. To further ensure that the distribution of the CN band was due to similar differences in underlying N abundances between RGB and AGB stars, we also compared the distribution of the CH band strengths to the RGB stars since it is independent of the N abundance.

\begin{figure}[htbp]
\centering
\includegraphics[trim = 0.7cm 0.4cm 0.4cm 0.4cm, scale=0.35, clip=True]{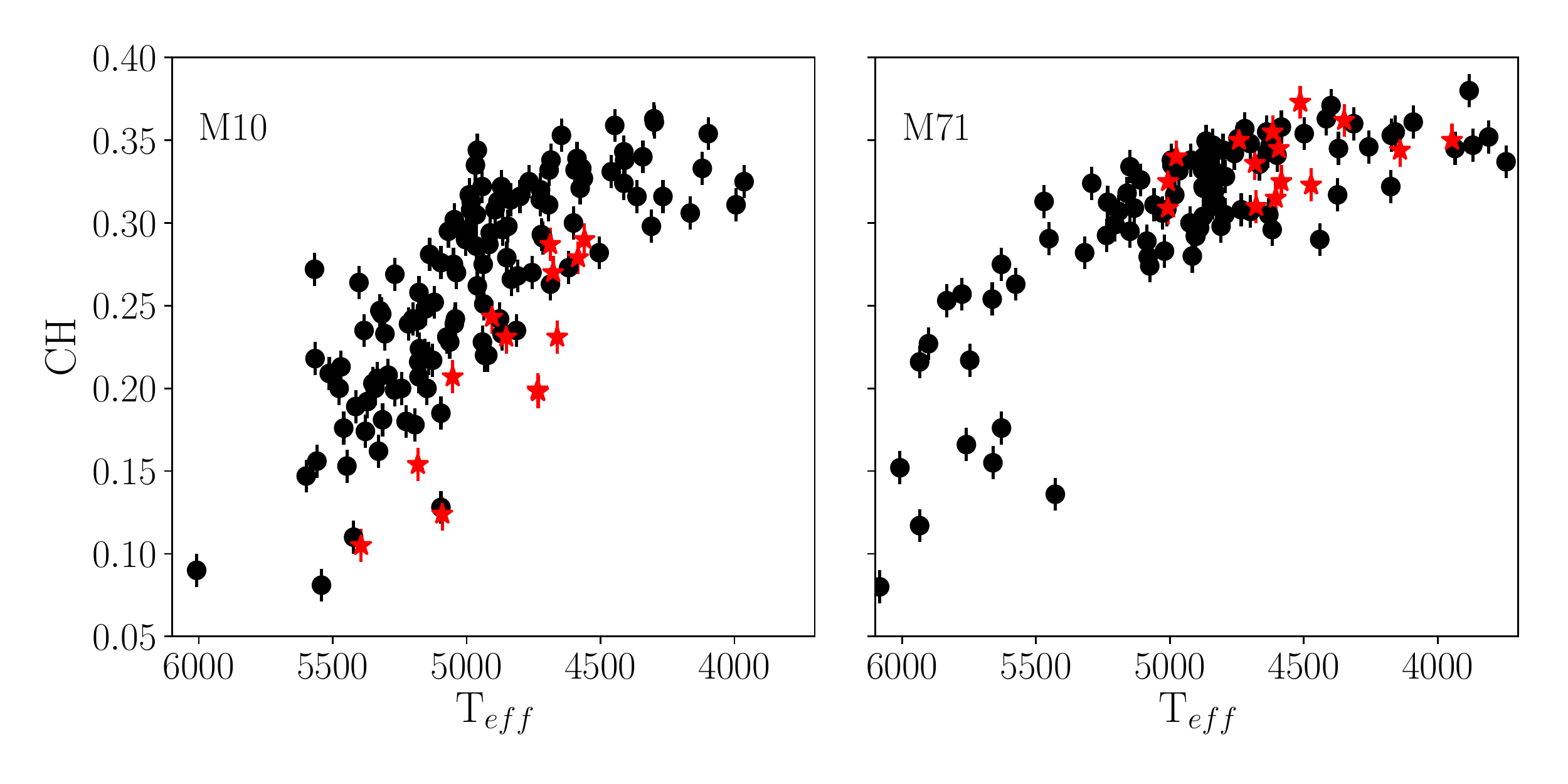}
\caption{CH(4300) band strength against effective temperature. Our measurements for M10 from G18 are shown on the left panel, and the results from this work for M71 are on the right. RGB stars are shown as black points and AGB stars are shown as red stars.}
\label{fig::chbandscomp}
\end{figure}

The right panel of Figure \ref{fig::chbandscomp} shows the CH band measurements for just the AGB and RGB stars in M71. The CH bands for the AGB stars match well with the CH bands for the RGB stars with similar T$_{eff}$, which is expected due to the less efficient extra mixing in a cluster of this metallicity. For comparison, we also show the CH band measurements from G18 from the RGB and AGB stars in M10 in the left panel of Figure \ref{fig::chbandscomp}. The AGB stars in this cluster have lower CH band measurements compared to RGB stars with similar T$_{eff}$ because the extra mixing is more efficient at the lower metallicity of M10 and causes stars to have lower surface carbon abundances than RGB stars by the time they reach the AGB phase. Thus the CH band can be used as an indication of whether or not the CN band can be used to securely sort AGB stars into populations. In the case of M71, we determine that since the CH bands of the AGB stars match the RGB stars, the separation in $\delta$CN index values between the two populations appearing in the AGB star distribution is caused by differences in N and not differences in C.

When using the same classification scheme as used for the RGB stars based on $\delta$S(3839), we find 6 CN-enhanced and 9 CN-normal AGB stars. If AGB stars are classified with the $\delta$CN(4216) index, the same distribution is found. This result equates to 40 $\pm$ 13\% of AGB stars sampled being CN-enhanced, which agrees well with the distribution found in the RGB stars of 42 $\pm$ 4\%. For M71, as in M10, it appears that there is no lack of the second generation stars on the AGB (G18).

\subsubsection{HB Stars}
Because the HB of M71 is relatively cool and red, we were able to obtain CN and CH measurements for 27 HB stars. We calculated $\delta$S(3839) and $\delta$CN(4216) indices for all HB stars using the same method as described in Section \ref{sec::bands} for the RGB stars. Our final values for each $\delta$CN index for the HB stars are plotted in Figure \ref{fig::deltacnall} as blue points. From Figure \ref{fig::deltacnall}, it appears that the HB stars separate into two populations similar to the RGB stars based on $\delta$CN(4216) band strength where two peaks can be seen in the distribution shown on the y-axis. However, while there is a spread in the $\delta$S(3839) index among HB stars, the distribution does not easily separate into two groups. To make sure that there are indeed two populations in the HB stars, we turned to the CH band measurements to see what information could be added.

\begin{figure}[htbp]
\centering
\includegraphics[trim = 0.2cm 0.4cm 0.4cm 0.4cm, scale=0.4, clip=True]{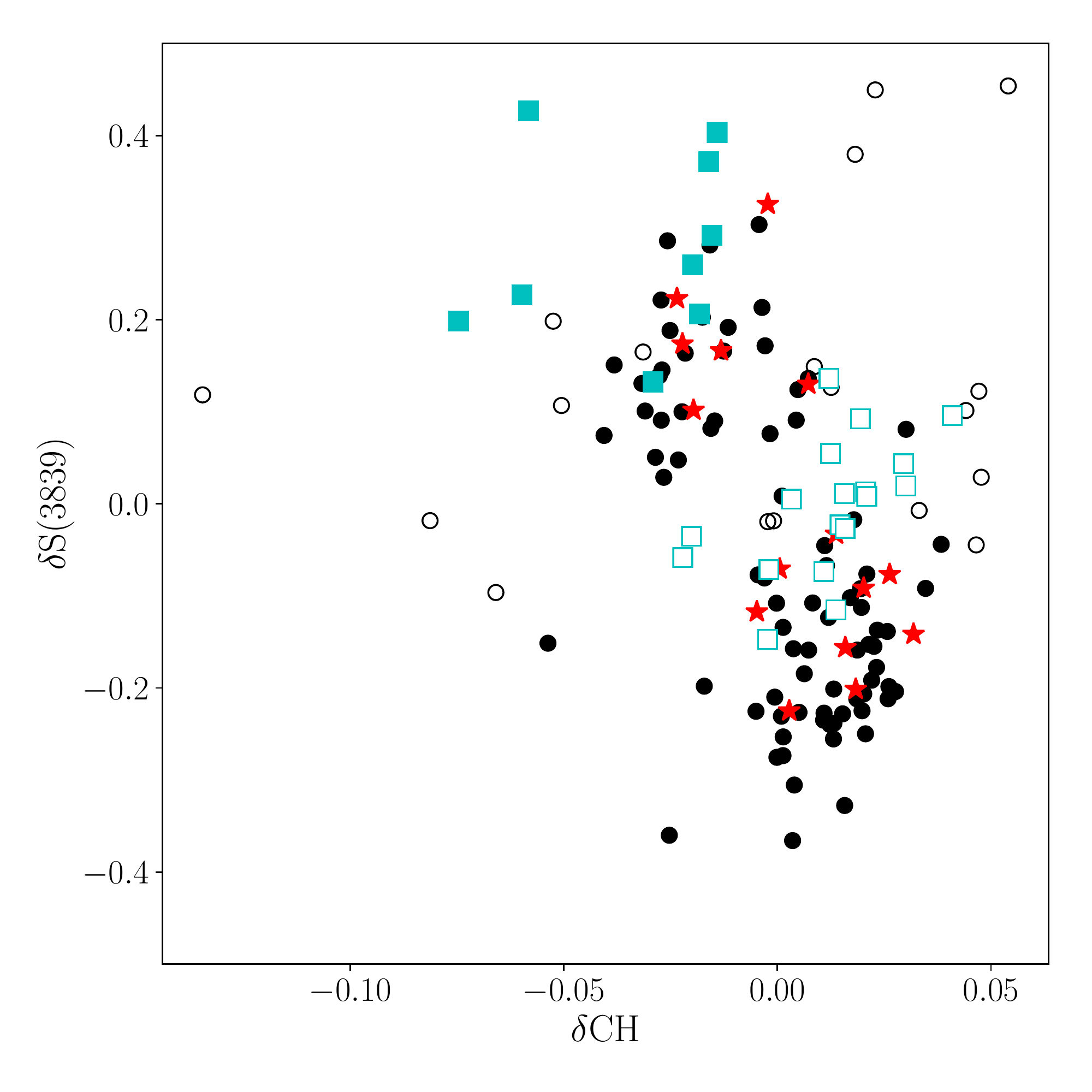}
\caption{$\delta$S(3839) vs. $\delta$CH for all stars in our sample. RGB, AGB, and HB stars are indicated as in Figure \ref{fig::cmd}. RGB stars with T$_{eff} >$ 5300 K are shown as open circles. HB CN-enhanced stars are cyan squares and CN-normal stars are open squares.}
\label{fig::chvscn}
\end{figure}

In Figure \ref{fig::chvscn}, we plot $\delta$S(3839) vs. $\delta$CH for all of the stars in our sample. The $\delta$CH index was created by fitting the CH band strength vs. T$_{eff}$ plotted in the bottom panel of Figure \ref{fig::bands} with a polynomial and subtracting this trend out of the CH band measurements. This method created a $\delta$CH measurement that was independent of temperature and surface gravity effects, and only depended on carbon abundance (similar to the $\delta$CN indices). Plotting the $\delta$S(3839) vs. $\delta$CH shows an anti-correlation between these two measurements as expected for RGB, AGB, and HB stars. The HB stars also separate in this plot into two populations, one CN-enhanced and CH-weak, and one CN-normal and CH-normal. This result indicates that although the distribution in $\delta$S(3839) alone does not show the same clear separation in populations as the $\delta$CN(4216) index, there are still two populations present in the HB stars given the evidence of two separated groups in the $\delta$S(3839) vs. $\delta$CH plot.

Because the HB stars separate into two groups more clearly in the $\delta$S(3839) vs. $\delta$CH plot (Figure \ref{fig::chvscn}), we used this diagram to classify the HB stars in our sample. We classified HB stars with $\delta$S(3839) values above the line $\delta$S(3839) = 4$\delta$CH + 0.2 as CN-enhanced and those below the line were classified as CN-normal. These stars are coded in Figure \ref{fig::chvscn} as open or closed squares for CN-normal and CN-enhanced, respectively. We find 9 CN-enhanced and 18 CN-normal stars on the HB, which gives 33 $\pm$ 9\% of stars as CN-enhanced. This value agrees with the RGB stars within our uncertainties, and provides evidence that there is not a lack of CN-enhanced stars on the HB in M71.

\begin{figure*}[htbp]
\centering
\includegraphics[trim = 0.2cm 0.4cm 0.4cm 0.4cm, scale=0.45, clip=True]{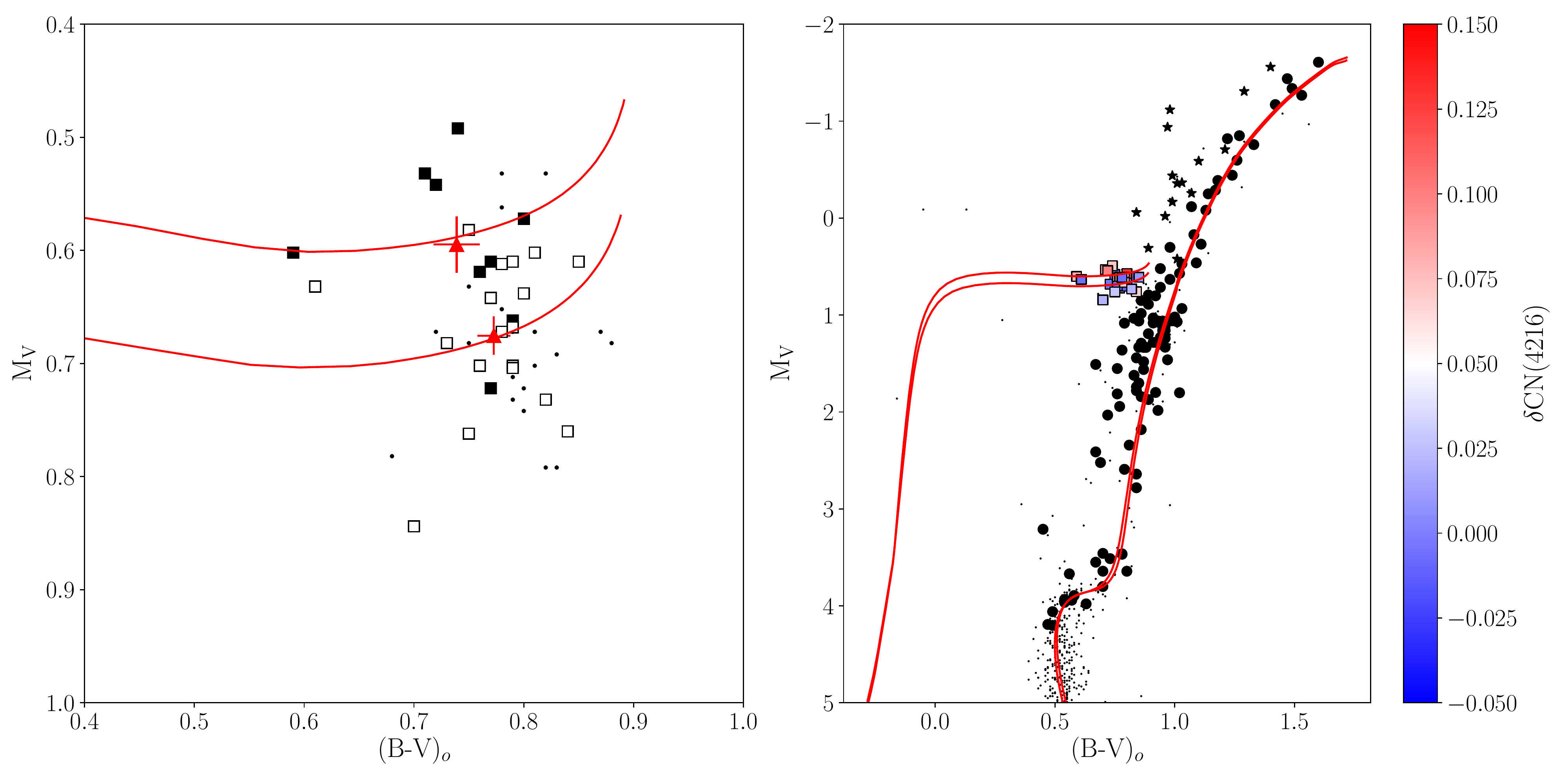}
\caption{\textbf{Left Panel:} CMD for M71 centered on the HB with M$_{\mathrm{V}}$ vs. (B-V)$_o$ (photometry from \citet{cudworth85}). Small dots indicate HB stars (determined based on their position in the CMD) that have an 85\% probability of membership determined by \citet{cudworth85} that were not measured for CN strength. HB stars are indicated as squares with open squares being CN-normal and filled being CN-enhanced following the same classification as shown in Figure 6. Average positions of each population are shown as red triangles with error bars indicating the mean errors. Red lines show the two ZAHBs that best fit these positions. \textbf{Right Panel} CMD of M71 with HB stars indicated as colored squares color-coded based on their $\delta$CN(4216) band strength as indicated by the color bar to the right. Black circles are RGB stars and stars are AGB stars. Overplotted are red lines showing the ZAHBs from the left panel as well as isochrones generated with the same parameters using the PGPUC isochrone interpolator.}
\label{fig::hbcmd}
\end{figure*}

A study of the metal rich GC, 47 Tuc, found that the CN-enhanced and CN-normal populations had different HB morphologies in the CMD \citep{briley97}. Since 47 Tuc has a metallicity similar to M71, with [Fe/H] = -0.72 (\citealt{harris} (2010 edition)), we analyzed the location of the two populations in M71 on the HB to see if the same difference in morphology was observed. The left panel of Figure \ref{fig::hbcmd} shows a zoomed in look of the CMD from Figure \ref{fig::cmd} centered on the HB with CN-enhanced stars shown as closed squares and CN-normal stars shown as open squares. We also show as red triangles the average Mv and (B-V)$_0$ of each population, with their associated mean errors. Although the scatter among individual stars is large, the CN-enhanced HB stars are displaced systematically to brighter magnitudes and bluer colors than the CN-normal HB stars, by differences that are large compared to the mean error of the average values (see Table \ref{tab::hbstars}).  
 This result is similar to what \citet{briley97} found in 47 Tuc, where the locus of the CN-enhanced population was found to be 0.05 magnitudes brighter than the CN-normal population; for M71 we find a difference of 0.08 mag.

As \citet{briley97} noted, a slight difference in He abundance between the two populations would cause a different zero-age horizontal branch location for each population. Multiple studies have now shown that the multiple populations observed in GCs have different He abundances \citep[see e.g.][, etc.]{milone15,milone18,lagioia18}. Higher envelope He content in second generation stars would cause these stars to sit at slightly bluer colors and slightly higher luminosities on the HB, just as observed in M71. To quantify the difference in He abundance between the two populations, we generated two zero-age horizontal branches (ZAHBs) with differing He abundances running through the average HB position of each populations. These ZAHBs were generated with the PGPUC isochrones, using the online interpolater that can be found at http://www2.astro.puc.cl/pgpuc/index.php \citep{valcarce12,valcarce13}, and are shown in Figure \ref{fig::hbcmd}. The right panel of Figure \ref{fig::hbcmd} shows the full CMD and isochrones generated with the PGPUC isochrone interpolator using the same parameters for the ZAHBs.

We found that the average position of the CN-enhanced stars is best fit by a ZAHB with Y = 0.325 and that of the CN-normal stars is best fit by a ZAHB with Y = 0.30. This difference of $\Delta$Y of 0.025 between the two populations agrees with the maximum $\Delta$Y found by \citet{milone18} using HST UV photometry, and confirms that the difference in average position on the HB between the two populations is likely caused by a difference in He abundance.

\begin{deluxetable}{ccc}
\tabletypesize{\small}
\tablecolumns{3}
\tablewidth{0pt}
\tablecaption{Mean HB Positions\label{tab::hbstars}}
\tablehead{\\
   \colhead{Population} &          \colhead{$M_\mathrm{V}$} &        \colhead{(B-V)$_o$}}
\startdata
CN-enhanced& 0.595 $\pm$ 0.07 ($\sigma$)& 0.739 $\pm$ 0.06 ($\sigma$)\\
&~~~~~~~~~~~~~$\pm$ 0.025 (m.e.)&~~~~~~~~~~~~~$\pm$ 0.021 (m.e.)\\
CN-normal& 0.675 $\pm$ 0.07 ($\sigma$)& 0.773 $\pm$ 0.06 ($\sigma$)\\
&~~~~~~~~~~~~~$\pm$ 0.017 (m.e.)&~~~~~~~~~~~~~$\pm$ 0.015 (m.e.)\\
\hline
$\Delta$& 0.08 & 0.034\\
\enddata
\end{deluxetable}

\subsection{Distribution Across the Cluster}
\begin{figure}[htbp]
\centering
\includegraphics[trim = 0.2cm 0.4cm 0.4cm 0.4cm, scale=0.4, clip=True]{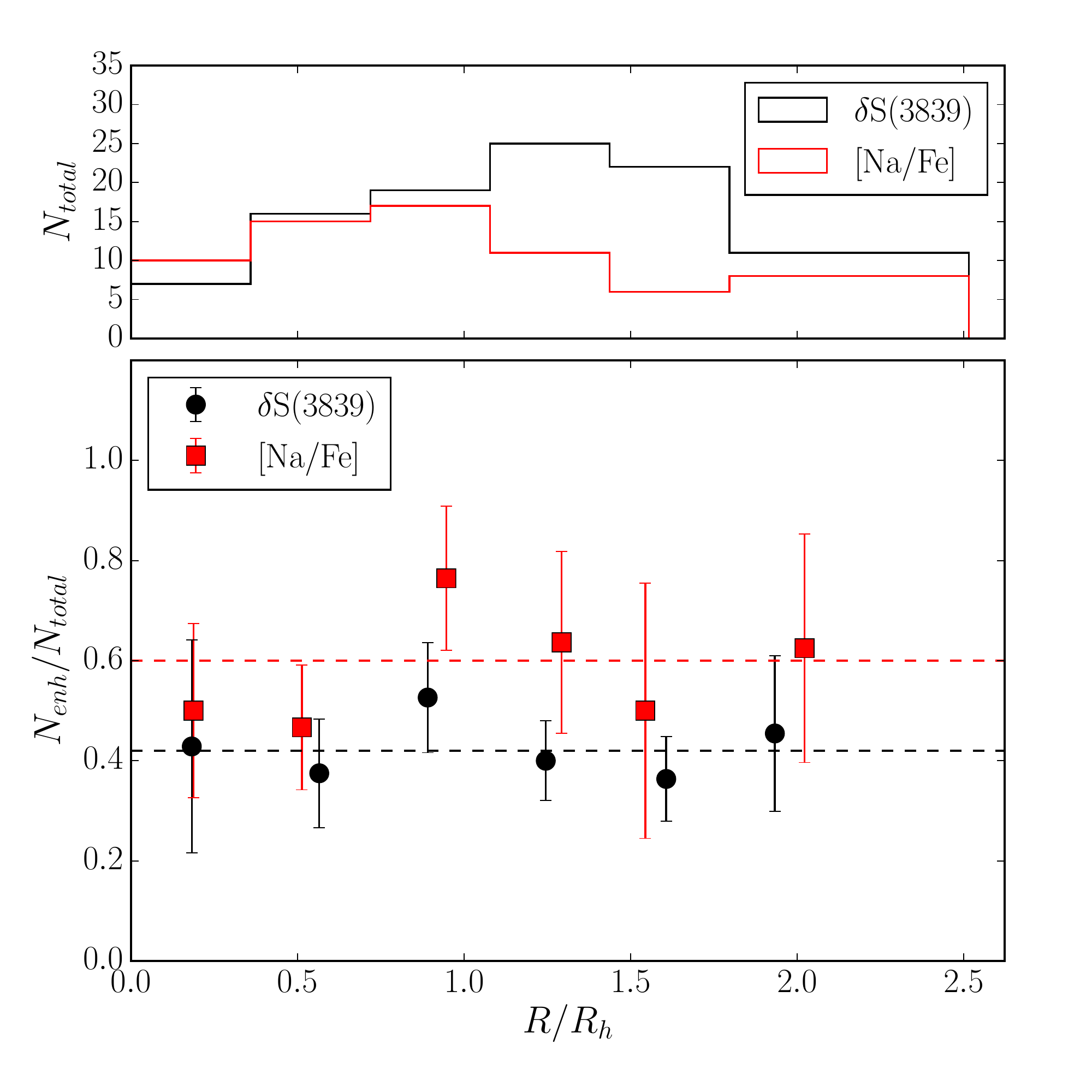}
\caption{\textbf{Top:} A histogram of RGB stars as a function of radial distance from the cluster center in units of half-light radius where R$_h$ = 100.2" (\citealt{harris} (2010 edition)). Our sample is shown in black, and RGB stars with a [Na/Fe] measurement from \citet{sneden94}, \citet{ramirez02}, \citet{carretta09a,carretta09b}, or \citet{cordero15} are combined into one sample shown in red. \textbf{Bottom:} The ratio of CN/Na-enhanced RGB stars to the total number of RGB stars in each radial bin from the top panel. Each point is at the average location of the stars in a given bin. The ratio calculated using the $\delta$S(3839) band to sort into populations is shown as black circles, and the ratio calculated using [Na/Fe] abundance is shown as red squares. Error bars are uncertainties based on Poisson statistics. A black dashed line is shown to indicate the ratio calculated for our entire sample. A red dashed lined is shown to represent the approximate ratio determined using the Na abundances from the literature and [Na/Fe] = 0.3 dex (from \citet{carretta09a,carretta09b}) as a separating value.}
\label{fig::radialdist}
\end{figure}

Thanks to the large number of stars observed and the broad radial coverage extending to $\sim$2.5 half-light radii (R$_h$ = 1.67 arcmin; \citealt{harris}, 2010 edition), we are able to analyze the radial distribution of the two populations found in M71. The bottom panel of Figure \ref{fig::radialdist} shows the ratio of the number of CN-enhanced to the total number of RGB stars in radial bins with a width of $\sim$0.35 half-light radii. The top panel of Figure \ref{fig::radialdist} shows a histogram of the distribution of stars used to calculate these ratios. The distribution from our sample using the $\delta$S(3839) index is shown as a black line in the top panel and as closed circles in the bottom panel. The distribution from a collection of Na abundance measurements from \citet{sneden94}, \citet{ramirez02}, \citet{carretta09a,carretta09b}, and \citet{cordero15} is shown as a red line in the top panel and as red squares in the bottom panel. The offsets we used to combine these samples come from \citet{smith15b} and are discussed further in Section \ref{sec::na-o}.

Based on the results shown in Figure \ref{fig::radialdist}, we found no significant change in the number ratio of CN-enhanced to CN-normal stars as a function of the radial distance from the cluster's center. We also compare this result to what is found if the stars are divided into first and second generation based on Na abundance. We decided to follow \citet{carretta09a,carretta09b} and use [Na/Fe] = 0.3 dex to separate stars based on Na abundance instead of [Na/Fe] = 0.2 dex, which was used by \citet{cordero15}. Figure \ref{fig::radialdist} shows that while the ratio for Na-enhanced to Na-normal is different from that found in our study (see section \ref{sec::rgbpops}), when populations are identified using Na there is no significant radial variation in the ratio of Na-enhanced to Na-normal. \citet{cordero15} and \citet{bowman17} have studied the radial distribution of the two populations in M71 using Na abundance and narrow band photometry centered on the S(3839) band, respectively, and both studies agree with our result that there is no radial change in the number ratio of the two populations.

While all formation scenarios for multiple populations agree that second generation stars are originally more centrally concentrated \citep[see e.g.][]{decressin07a,decressin07b,dercole08,bekki10}, the results of Figure \ref{fig::radialdist} are not surprising considering that M71 is a dynamically old cluster. As \citet{cordero15} and \citet{bowman17} have pointed out, $t/t_{rh}\sim 44$ for M71 considering a cluster age of 12 Gyr \citep{dicecco15} with $t_{rh}$ being the half-mass relaxation time. For comparison, 47 Tuc, a cluster with a similar age and metallicity, has a much younger dynamical age of $t/t_{rh}\sim 3$ (using $t_{rh}$ value from \citealt{harris}, 2010 edition) and is observed to have a more centrally concentrated second generation compared to the first generation \citep{milone12,cordero15}. Other examples of clusters with more concentrated second generation stars can be found in \citet{bellini09}, \citet{lardo11}, \citet{carretta10}, \citet{johnson12}, \citet{milone12}, and \citet{simioni16}. Clusters in which second generation stars are still more concentrated than first generation stars tend to be, in general, dynamically younger \citep[see e.g.][for a recent observational study of the dependence of  the degree of spatial mixing on clusters' dynamical age]{dalessandro19}.

However, the simulations from \cite{vesperini13} have shown that as clusters experience the effects of internal relaxation and mass loss, these radial gradients in the populations' number ratio are gradually erased and dynamically old clusters may reach complete spatial mixing. In particular, \citet{vesperini13} found that to reach complete spatial mixing as observed in M71, M10 (G18), and a variety of other clusters \citep[see also][]{dalessandro14,nardiello15}, a cluster must have lost 60-70\% of its initial mass due to two-body relaxation (see also \citealt{miholics15}). Evidence for M71 having experienced enough mass loss to reach complete spatial mixing comes from multiple sources such as the flattened stellar mass function \citep{demarchi07}. \citet{carretta10} have suggested that the excess x-ray sources found in the cluster by \cite{elsner08} also imply that the cluster has experienced significant mass loss. Finally, a cluster's mass loss rate is enhanced by the stronger tidal field of the inner regions of the Galaxy \citep[see e.g][]{vesperini97,baumgardt03}. \citet{dinescu99} determined the orbit of M71 is confined within 6.7 kpc, and according to the recent orbital parameters' determination based on Gaia data \citep{helmi18}, the pericenter of M71's orbit is at about 5 kpc from the Galactic center and its apocenter at about 7.2 kpc. These results suggest that the cluster is indeed likely to have experienced significant mass loss.

\subsection{Comparison to Other Definitions of Multiple Populations}
\subsubsection{Na-O Anti-Correlation} \label{sec::na-o}
Multiple populations in GC's have differences in their Na and O abundances as well as their C and N abundances. Due to the various nucleosynthesis processes in the progenitors of the material that forms the second generation of stars, the second generation should be enhanced in Na and depleted in O compared to the first \citep{gratton12}. The two main processes involved that create these anti-correlations (C-N and Na-O) are the CN(O)-cycle and the Ne-Na cycle. The C-N anti-correlation is a direct result of the CN(O)-cycle as it works in the progenitor stars to turn carbon into nitrogen. This carbon-depleted and nitrogen-enriched material is then ejected from the progenitor stars and enriches the intra-cluster medium, which then forms the second generation. The Na-O correlation is a result of both cycles working alongside each other. As the CN(O)-cycle causes a depletion in oxygen, the Ne-Na cycle converts neon to sodium in the same progenitor stars. Once again, this material is ejected from the progenitors and pollutes the intra-cluster medium, which the second population forms from. The second generation is therefore depleted in carbon and oxygen, and enhanced in nitrogen and sodium compared to the first.

\begin{figure*}[htbp]
\centering
\includegraphics[trim = 0.2cm 0.4cm 0.4cm 0.4cm, scale=0.55, clip=True]{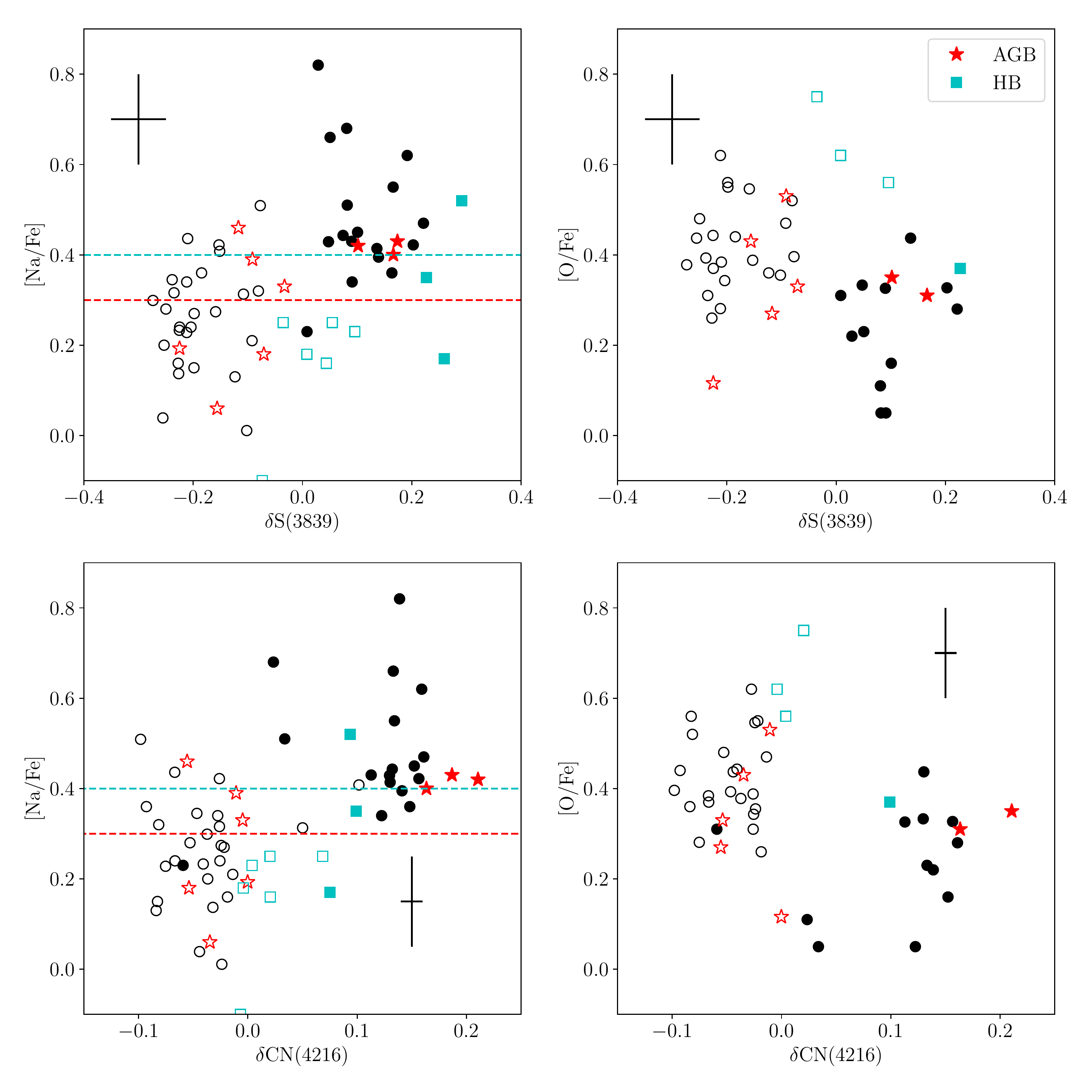}
\caption{\textbf{Top Left:} [Na/Fe] vs. $\delta$S(3839) for all stars in our sample with [Na/Fe] measurements. A red dashed line is placed where the division in populations is defined for [Na/Fe] \citep{carretta09a,carretta09b}. A blue dashed line shows the separation in [Na/Fe] increased by 0.1 dex. \textbf{Top Right:} [O/Fe] vs. $\delta$S(3839) for all stars in our sample with [O/Fe] measurements. \textbf{Bottom Row}: The same as the top row except for $\delta$CN(4216) instead of $\delta$S(3839). In all panels, symbols are coded as in Figure \ref{fig::cmd}, CN-enhanced stars are filled, and CN-normal stars are open.}
\label{fig::deltacnvsna}
\end{figure*}

This Na-O anti-correlation has been used frequently to identify multiple populations in GCs \citep[see e.g.][and references therein]{carretta09a,carretta09b,gratton12}. Multiple studies have looked at how N correlates with Na and O to see if these elements are classifying stars into the same populations \citep[see e.g.,][and references therein]{smith13,smith15a,smith15b,bobergm53}. Any outliers among these correlations could hint at stars with an interesting nucleosynthesis history and provide additional constraints for scenerios of multiple populations' formation. 
While most stars in the GC's studied fell into the same population regardless of whether CN band strength or Na abundance was used, a few stars in M5, 47 Tuc, and M53 have been found to have CN-enhanced band strengths without having enhanced Na abundances \citep{smith13,smith15a,smith15b,bobergm53}.

As we stated in Section \ref{sec::observations}, we prioritized observing stars with Na and O measurements in M71 to compare these two methods of identifying multiple populations with a large homogeneous sample. Final Na and O abundances for our stars come from \citet{sneden94}, \citet{ramirez02}, \citet{carretta09a,carretta09b}, and \citet{cordero15}. \citet{smith15b}, in his analysis of M71, determined the small offsets in abundances measured from these sources ($\sim$0.1 to 0.2 dex). We adopt the same offsets to put all samples cited above on the same standard scale for our study.

Since we have two CN band measurements for our stars in M71, we compare both of them to the Na and O abundances, which can be seen in Figure \ref{fig::deltacnvsna} where we have plotted [Na/Fe] and [O/Fe] vs. $\delta$S(3839) and $\delta$CN(4216). Overall, we find that both of our $\delta$CN indices correlate with [Na/Fe] and anti-correlate with [O/Fe] as expected for RGB, AGB, and HB stars, even though the HB stars seem to have slightly lower Na and slightly higher O than the RGB stars in the sample. To better evaluate whether there are any CN-normal stars with enhanced Na or depleted O or vice versa, we have indicated the population of each star as classified by this work in Figure \ref{fig::deltacnvsna} by representing CN-enhanced stars as filled points and CN-normal stars as open. Once stars are coded in this way, Figure \ref{fig::deltacnvsna} shows that most stars fall in the correct place in the Na vs. $\delta$CN and O vs. $\delta$CN planes, save for a few exceptions. There is one HB star (ID = 153) that has been classified as CN-enhanced, but is found to have a [Na/Fe] value of $\sim$0.2 dex, which better matches the [Na/Fe] values of the CN-normal population. There is also one AGB star (ID = 102) that has very depleted O for a CN-normal star, but its [Na/Fe] value matches that of other CN-normal stars. As far as RGB stars that appear out of place, we identified two stars in Section \ref{sec::bands} that were misidentified by their $\delta$S(3839) measurements, and both of these appear out of place in Figure \ref{fig::deltacnvsna}.

The RGB star with ID = 63 (with $\delta$S(3839) = 0.008 and [Na/Fe] = 0.230) is one of the coolest stars in our sample (T$_{eff}$ = 3810 K), which means it was misidentified due to the S(3839) band losing sensitivity faster than the CN(4216) band at cooler temperatures. This star was identified by the $\delta$S(3839) band as a CN-enhanced star, and it appears in the [Na/Fe] vs. $\delta$S(3839) plot (top left panel of Figure \ref{fig::deltacnvsna}) as one of the CN-enhanced stars with a low [Na/Fe] value. However, from the [Na/Fe] vs. $\delta$CN(4216) plot and the [O/Fe] vs. $\delta$CN(4216) (bottom left and right panels of Figure \ref{fig::deltacnvsna}), it is clear that the star is misidentified by its $\delta$S(3839) as its [Na/Fe], [O/Fe], and $\delta$CN(4216) measurements all match the CN-normal population. Similarly, the other misidentified RGB star (ID = 99, $\delta$S(3839) = -0151, [Na/Fe] = 0.408), is a star that was identified as CN-normal, but from Figure \ref{fig::deltacnvsna}, it appears to have a [Na/Fe] and $\delta$CN(4216) index measurement similar to the CN-enhanced population. Fortunately, these misidentifications by $\delta$S(3839) do not affect our calculations for the ratio of CN-enhanced to CN-normal stars in Section \ref{sec::rgbpops} because we have one misidentified star in each population. There are also two CN-enhanced RGB stars with a $\delta$CN(4216) measurement close to 0.0 that have very low [O/Fe] measurements. These stars have lower $\delta$CN(4216) values than other CN-enhanced stars because they are two of the coolest stars in the sample (T$_{eff} \leq$ 3900), which means the CN(4216) band is not as sensitive to underlying changes in N abundance. They are likely still identified correctly as they have enhanced Na and depleted O as seen in Figure \ref{fig::deltacnvsna}.

Finally, Figure 9 shows some RGB stars that are CN-normal, but fall above the [Na/Fe] = 0.3 dividing line used by \citet{carretta09a,carretta09b} to denote the Na-enhanced population.  We note that while the bimodal distribution of $\delta$CN leads to a relatively secure assignment to populations, the continuous nature of the [Na/Fe] distribution makes such a separation difficult.  The identification of a single value of [Na/Fe] to classify different populations is likely to lead to the misclassification of a few stars near the chosen Na threshold. This can easily lead to some stars in each of the CN populations having slightly high or low Na values compared to the other stars in their population.

As \citet{carretta09a,carretta09b} point out, the specific value of [Na/Fe] adopted in their analysis to separate the first and second generation stars is just indicative and not meant to provide a strict classification.  We note that the CN-normal stars above the Na dividing line are at most only $\sim$0.2 dex more enhanced than other CN-normal stars, and that they no longer seem out of place if the line separating the populations in [Na/Fe] is moved up just 0.1 dex to [Na/Fe] = 0.4 dex, which is shown in the top and bottom left panels of Figure \ref{fig::deltacnvsna} as a blue dashed line.  Setting the separation to [Na/Fe] = 0.4 dex also changes the percentage of stars identified to be second generation stars by Na abundance from 60\% to 33\%, which comes closer to the value we find based on our $\delta$CN strengths from Section \ref{sec::rgbpops} and that found by \citep{bowman17}.

\subsubsection{HST UV Legacy Archive Photometry} \label{sec::hst}
HST UV photometry has also been used in recent years to identify multiple populations in GC's based on a pseudocolor that is dependent on the C, N, and O abundances in a star \citep[see][]{piotto15,milone17}. Designed as a way to increase the separation between the two populations seen in RGB's of the CMD of a cluster, this pseudocolor is made up of HST filters that are centered on NH, CH, OH, and CN bands. Since these bands are affected by C, N, and O abundances, which are different between the populations in GC's, the different populations separate themselves well enough in the pseudo-CMD to be identified with N-enhanced stars appearing on the ``bluer'' branch and N-normal stars appearing on the ``redder'' branch.

\begin{figure}[htbp]
\centering
\includegraphics[trim = 0.2cm 0.4cm 0.4cm 0.4cm, scale=0.4, clip=True]{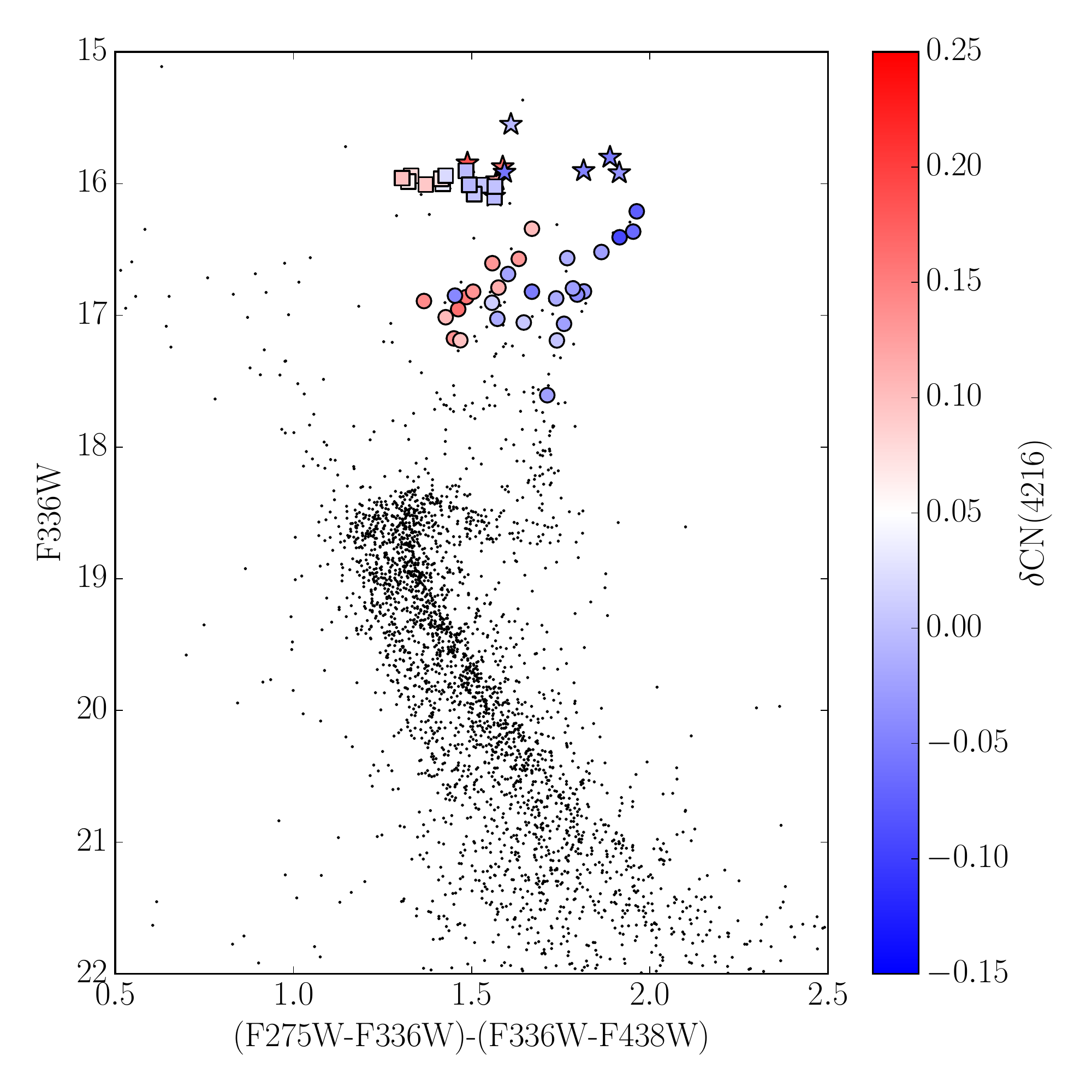}
\caption{HST UV photometry for M71 using the pseudo-CMD as seen in \citet{milone17}. Stars from our sample are indicated with RGB stars as circles, AGB stars as stars, and HB stars as squares. All stars in common with this study are color-coded based on $\delta$CN(4216) value.}
\label{fig::hstcmd}
\end{figure}

M71 was included in the HST UV Legacy Survey of Galactic Globular Clusters conducted by \citet{milone17}, and this photometry is shown in Figure \ref{fig::hstcmd}. M71 has a clear separation on the RGB into two populations, and \citet{milone17} find that 37.8 $\pm$ 3.8\% of stars in the cluster belong to the second generation, in agreement with our findings based on $\delta$CN measurements.

In Figure \ref{fig::hstcmd} we also show stars in common between the HST photometry and our study color-coded based on $\delta$CN(4216) strength. RGB stars are indicated as circles, AGB stars as stars, and HB stars as squares. For all three evolutionary stages of stars, CN-enhanced stars and CN-normal stars generally fall in the correct place on the pseudo-CMD with CN-enhanced stars being ``bluer'' than CN-normal stars. There is one RGB star that is CN-normal but sits on the ``blue'' CN-enhanced star sequence. A possible explanation for this is the relatively high amount of reddening seen towards the direction of this cluster compared to other GC's. This reddening would have a larger effect on the bluer photometric bands used for this analysis, and has also been shown to change across the cluster \citep{morrison16}. Variable reddening could possibly cause a star to appear ``redder'' than it actually is. Even with this one outlier, we are left with the result that the HST UV photometry method and our $\delta$CN method of identifying multiple populations are classifying stars into the same populations in general, which agrees with what we found in M10 (G18).

\section{Conclusions}
We have classified 100 RGB, 15 AGB, and 27 HB stars in M71 into CN-enhanced and CN-normal populations based on CN band measurements made with low-resolution spectroscopy. We then calculate the ratio of first and second generation stars in each evolutionary stage, find the He abundance difference between the populations, study the radial distribution of each population, and compare our methods of classification to others used in the literature to identify multiple populations in GCs. Our conclusions based on this analysis are listed below:

\begin{enumerate}
\item We classify stars into populations based on both the S(3839) and CN(4216) CN band indices measured for this cluster. For the brighter, cooler stars (M$_{\mathrm{V}} \leq$  3.0, T$_{eff} \leq$  5300 K), the separation in the CN(4216) band strength can be clearly seen by eye. We create a $\delta$S(3839) and $\delta$CN(4216) index and fit both distributions with Gaussian mixture models. These models indicate that two populations are needed to best explain the distribution of both $\delta$CN indices. We find that unlike most clusters, which are dominated by the CN-enhanced population \citep[see e.g.][]{milone17}, M71 is a CN-normal, first generation dominated cluster with 58 $\pm$ 4\% of RGB stars belonging to the first generation.

\item We determine a C and N abundance for each population based on matching $\delta$S(3839) and $\delta$CN(4216) indices vs. effective temperature with models generated with the SSG. The CN-enhanced population is best fit with [C/Fe] = -0.3 dex and [N/Fe] = 1.5 dex, and the CN-normal population is best fit with [C/Fe] = 0.0 dex and [N/Fe] = 0.6 dex. These results agree well with what was found for giant stars from DDO photometry by \citet{briley01b} and with C and N abundances measured for main sequence stars by \citet{briley01a,briley04b}. The separation in [N/Fe] between the two populations of 0.9 dex is comparable to the 0.7 dex separation found in M10 (G18).

\item The AGB and HB stars in M71 separate into two populations with a similar ratio between CN-enhanced and CN-normal stars as the RGB stars. This means that both generations of stars are moving onto the HB and AGB. Furthermore, the HB stars' location on the CMD seems to be dependent on CN band strength with the average position of the CN-enhanced stars on the HB being brighter than the average position of the CN-normal stars by 0.08 magnitudes. This result is consistent with what is seen in 47 Tuc, a cluster of similar metallicity \citep{briley97}. Through ZAHB fitting using PGPUC isochrones, we determine that this difference in HB position is consistent with the CN-enhanced stars having a Y value higher by 0.025. This $\Delta$Y is consistent with what is found by \citet{milone18} using HST UV photometry.

\item We find that the fraction of CN-enhanced stars does not vary with the distance from the cluster's center. The same result is found when [Na/Fe] is used to classify stars. M71 is a dynamically old cluster and the complete spatial mixing of the two populations observed is consistent with being the result of dynamical evolution.

\item Based on stars in our sample with Na and O measurements \citep{sneden94,ramirez02,carretta09a,carretta09b,cordero15} and those with HST UV photometry \citep{milone17}, we find no major discrepancies between any of the different methods of classifying multiple populations in M71. While the HB stars seem to have slighly lower Na abundances than the RGB stars, they still correlate with $\delta$CN strength as expected. There is a group of stars that appear to be CN-normal with slightly enhanced Na based on the definition separating populations from \citet{carretta09a,carretta09b} of [Na/Fe] = 0.3 dex, but if the separating value is raised by only 0.1 dex to [Na/Fe] = 0.4 dex, then this group no longer exists. Raising the separating value also brings the ratio of second generation stars from 60 \% to 33 \%, which comes closer to agreement with the ratio we find using $\delta$CN, the ratio found by \citet{milone17} with HST photometry, and the ratio found by \citet{bowman17} with narrow, CN-band photometry. Additionally with the exception of one HB star, we do not observe any CN-enhanced stars with normal [Na/Fe] values as reported by \citet{smith13}, \citet{smith15a}, and \citet{bobergm53} in M5, 47 Tuc, and M53, respectively.
\end{enumerate}

\section{Acknowledgements}
We would like to thank Roger A. Bell for making the SSG program available to us, and Kyle Cudworth for sharing his M71 proper motions and membership probabilities. We would also like to thank Zachary Maas for his help in obtaining the stellar spectra during our 2016 run at WIYN. Additionally, we would like to thank Dianne Harmer, Daryl Willmarth, and all of the observing assistants who helped with our multiple runs on Hydra. 

This publication makes use of data products from the Two Micron All Sky Survey, which is a joint project of the University of Massachusetts and the Infrared Processing and Analysis Center/California Institute of Technology, funded by the National Aeronautics and Space Administration and the National Science Foundation.

\end{document}